%% file: main.tex
\documentclass[
  journal=pasa,
  manuscript=research-paper, 
  year=202X,
  volume=YY,
]{cup-journal}

\usepackage[T1]{fontenc}
\usepackage{upgreek}
\usepackage{amsmath}
\usepackage{amssymb,siunitx,booktabs}
\usepackage{graphicx}	
\usepackage{amsmath}	
\usepackage{siunitx}    
\usepackage{enumitem}
\usepackage{soul}
\usepackage{subcaption}
\usepackage{tabularx}
\usepackage{bm}
\usepackage[table, dvipsnames]{xcolor}
\usepackage{caption}
\usepackage{hyperref}

\newcommand{\ha}{\ensuremath{{\rm H}\upalpha}}
\newcommand{\hb}{\ensuremath{{\rm H}\upbeta}}

\newcommand{\hii}{H\textsc{ii}}

\newcommand{\oiii}{[O\textsc{iii}]}
\newcommand{\nii}{[N\textsc{ii}]}
\newcommand{\ppxf}{\textsc{pPXF}}
\newcommand{\lzifu}{\textsc{lzifu}}

\newcommand{\sigmastar}{\ensuremath{\sigma_*}}

\newcommand{\kms}{\ensuremath{\rm \,km\,s^{-1}}}
\newcommand{\msol}{\ensuremath{\rm \,M_\odot}}
\newcommand{\re}{\ensuremath{R_{\rm e}}}
\newcommand{\lothree}{\ensuremath{L\mathrm{[O\textsc{iii}]}}}
\newcommand{\los}{\ensuremath{L\mathrm{[O\textsc{iii}]}/\sigma_*^4}}
\DeclareSIUnit\angstrom{\text {Å}}

\setlist[enumerate]{labelwidth=0pt}

\title[New Techniques for AGN-SF Study with IFU Data]{New techniques to investigate the AGN-SF connection with integral field spectroscopy}

\author{Aman Chopra}
\affiliation{Australian National University, Canberra, ACT 2611, Australia; ARC Centre of Excellence for All Sky Astrophysics in 3 Dimensions (ASTRO 3D)}
\email[Aman Chopra]{aman@mso.anu.edu.au}

\author{Henry R. M. Zovaro}
\affiliation{Australian National University, Canberra, ACT 2611, Australia; ARC Centre of Excellence for All Sky Astrophysics in 3 Dimensions (ASTRO 3D)}

\author{Rebecca L. Davies}
\affiliation{Swinburne University of Technology, Hawthorn, VIC 3122, Australia; ARC Centre of Excellence for All Sky Astrophysics in 3 Dimensions (ASTRO 3D)}

\received{26 Sep 2025}
\revised  {24 Dec 2025}
\accepted {16 Jan 2026}
\published{dd Mmm 2026}


\begin{document}

\begin{abstract}
Understanding the connection between active galactic nuclei and star-formation (the AGN-SF connection) is one of the longest standing problems in modern astrophysics. In the age of large Integral Field Unit (IFU) surveys, studies of the AGN-SF connection greatly benefit from spatially resolving AGN and SF contributions to study the two processes independently. Using IFU data for 54 local active galaxies from the S7 sample, we present a new method to separate emission from AGN activity and SF using mixing sequences observed in the [N\textsc{ii}]$\uplambda$6584$\si{\angstrom}$/\ha{} vs. [O\textsc{iii}]$\uplambda$5007$\si{\angstrom}$/\hb{} Baldwin-Phillips-Terlevich (BPT) diagram. We use the new decomposition method to calculate the \ha{} star-formation rate and AGN \oiii{} luminosity for the galaxies. Our new method is robust to outliers in the line ratio distribution and can be applied to large galaxy samples with little manual intervention. We infer star-formation histories (SFHs) using \ppxf, conducting detailed recovery tests to determine the quantities that can be considered robust. We test the correlation between the AGN Eddington ratio, using the proxy \los{}, and star-formation properties. We find a moderately strong correlation between the Eddington ratio and the star-formation rate (SFR). We also observe marginally significant correlations between the AGN Eddington ratio and the light-weighted stellar age under 100 Myr. Our results point to higher AGN accretion being associated with young nuclear star-formation under 100 Myr, consistent with timelines presented in previous studies. The correlations found in this paper are relatively weak; extending our methods to larger samples, including radio-quiet galaxies, will help better constrain the physical mechanisms and timescales of the AGN–SF connection.
\end{abstract}

\section{Introduction}
\label{sec:int}

The tight correlation between the masses of supermassive black holes (SMBHs) and the velocity dispersion of the stars in the bulge of their host galaxy suggests that the growth of supermassive black holes and galaxy bulges is connected \citep{ah2012, konho2013}. Cosmological simulations over-predict the number of massive galaxies when AGN feedback is not incorporated \citep{Bower2006, Croton2006}, suggesting that AGN `negative feedback' plays a major role in the regulation and quenching of star-formation in massive galaxies. AGN feedback can prevent gas from cooling by heating material in the interstellar medium through radiation \citep{ss2006}, and redistribute or expel cold gas from the central region of a galaxy by driving strong outflows on parsec to kiloparsec scales \citep{debuhretal2012}. Radio plasma jets are important drivers of AGN negative feedback \citep{morganti2013, alatalo2015}.

On the other hand, enhanced AGN accretion is associated with recent star-formation. \citet{silverman2009} found that the incidence of AGN activity is inversely proportional to the average stellar age. Several studies conclude that at least 30-50$\%$ of Seyfert 2 nuclei are associated with recent star-formation \citep[$<100$ Myr; e.g.,][]{ghl2001, fernandes2004, delgado2005}. 

Studies of the AGN-SF connection often use optical diagnostic diagrams (ODDs) to classify the ionizing sources present in galaxies. The most common ODD is the [N\textsc{ii}]$\uplambda$6584$\si{\angstrom}$/\ha{} vs. [O\textsc{iii}]$\uplambda$5007$\si{\angstrom}$/\hb{} diagram or the Baldwin-Phillips-Terlevich (BPT) diagram \citep{bpt, veilleuxosterbrock1987}. The BPT diagram separates \hii{} regions from objects photoionized by a harder radiation field, such as AGN or shocks. Below the empirical line derived by \citet{ka03} (hereafter Ka03), galaxies are dominated by star-formation (SF). Galaxies above the \citet{ke01} line (hereafter Ka01), derived from theoretical stellar photoionization models, must have a significant contribution from AGN, shocks, or hot low-mass evolved stars (HOLMES). Composite galaxies lie between the Ka03 and Ke01 lines \citep{ke2006} and exhibit spectra with significant contributions from a combination of SF, AGN, shocks, and HOLMES.

The right arm on the BPT diagram is known as the starburst-AGN mixing sequence~\citep{heckman04, kewley2006, ka2009}. Line ratios at the bottom of this sequence are characteristic of \hii{} regions and hence are dominated by star-formation. Points further along the mixing sequence have a larger relative contribution from harder ionizing sequences such as AGN or shocks which increase the rate of collisional excitation. Thus, to first order, the distance along the mixing sequence is a metric for the fractional contribution of AGN ($f_{\textrm{AGN}}$) to the emission line spectrum of a galaxy, in the absence of any contamination from non-AGN hard ionizing sources. 

\cite{wild2010} analysed a sample of 400 post-starburst bulge galaxies from the Sloan Digital Sky Survey (SDSS) to construct a time-averaged view of black hole accretion following strong circumnuclear bursts of star-formation. They computed AGN fractions and isolated the AGN contribution to the [O\textsc{iii}] line. They then converted the AGN [O\textsc{iii}] luminosity to AGN bolometric luminosity and black hole accretion rate (BHAR). They found a sharp rise in BHAR approximately 250 Myr after the onset of a starburst, which was consistent with a model where 0.5 per cent of the mass shed by bulge-residing, intermediate-mass stars is accreted onto the black hole. High-energy supernova feedback was proposed as a mechanism that could suppress the BHAR during the early phase of the starburst.

A significant limitation of the \cite{wild2010} study was that it used SDSS data, which was captured using a single-fibre spectrograph with a fixed aperture, covering approximately 1\,kpc on average within each galaxy. The conversion between [N\textsc{ii}]/\ha{} and [O\textsc{iii}]/\hb{} ratios to absolute AGN fractions depends on properties of the line-emitting gas and the AGN, including the metallicity and ionization parameter of the gas, and the hardness of the AGN radiation field which vary significantly between galaxies and throughout a given galaxy (see figure~3 in \citealt{Davies2016}; see also \citealt{Kewley2019} for a review). Therefore, AGN fractions calculated from aperture or galaxy-integrated line ratios have significant systematic uncertainties. 

With the advent of integral field spectroscopy, it is now possible to map stellar population properties, star-formation rates and the impact of AGN activity across galaxies \citep[e.g.,][]{davies2014a, mulcaheyetal2022}. \citet{fernandesetal2013} used IFU data from the Calar Alto Legacy Integral Field Area (CALIFA) survey, one of the first large optical IFU surveys imaging nearby galaxies with kiloparsec-scale spatial resolution \citep{sanchezetal2012}, to collapse multi-dimensional data across time, metallicity and 2D spatial coordinates into radius-age diagrams that trace how stellar light, stellar mass and star formation rate have shifted over time, providing evidence of inside-out galaxy growth. \citet{lacerdaetal2020} compared a sample of nearby active galaxies ($z < 0.1$) with their non-active counterparts from the extended CALIFA (eCALIFA) survey to study the quenching of star-formation. They found a "mixed scenario" whereby AGN feedback expels or heats molecular gas, and the growth of the bulge subsequently prevents further star formation by increasing the gas depletion time. 

Studies have also compared SFRs from ionized gas emission lines and stellar population fitting in AGN host galaxies using MaNGA~\citep{bundyetal2015}, a multiplexed IFU survey of $\sim \num{10000}$ nearby galaxies. \citet{riffeletal2021} accounted for AGN contamination by including an AGN power-law continuum in their stellar fitting, deriving a correction between SFR from \ha{} and stellar population fits over the last 20 Myr. Similarly, \citet{demellosetal2024} calibrated \ha{} and \oiii{} to estimate SFRs, finding good agreement with stellar population-based SFRs.

\citet{Davies2016} calculated AGN fractions for two galaxies from the Siding Spring Southern Seyfert Spectroscopic Snapshot Survey (S7). The method involved expressing emission line fluxes in each spaxel (the `spatial pixel' associated with a spectrum in a reconstructed IFU datacube) as weighted combinations of SF and AGN `basis spectra' representing the lower and upper ends of the mixing sequences in the BPT diagram. The basis spectra were determined by fitting a line to the mixing sequence in emission line ratio space. The results of this method were consistent with high-resolution imaging tracing the star-forming and AGN regions, demonstrating the success of the decomposition. However, this method for determining basis spectra has weaknesses. The fitting has stringent requirements for the mixing sequence to have sufficiently low scatter and requires manual tweaking, precluding its application to a large number of galaxies.

We build upon \citet{Davies2016} and previous work by introducing a novel method of calculating basis spectra which is more robust to outliers and mixing sequences of varying shapes. We apply this method to high spatial resolution IFU data of 54 galaxies from the S7 survey to spatially resolve AGN and SF contributions. This method results in more accurate star-formation rates and AGN accretion rates than those derived from single-aperture data. 

To recover the star-formation histories (SFHs) of the AGN host galaxies, we fit the continuum emission using the Penalised PiXel Fitting method~\citep[\ppxf{};][]{Cappellari&Emsellem2004,Cappellari2017}. We run a series of tests using mock spectra to investigate the impact of factors such as the contamination from the AGN continuum and extinction on the derived SFH from \ppxf{} to ensure that our stellar ages are reliable. We analyse correlations between the AGN Eddington ratio and SF properties and recover weak but consistent trends, paving the way for future AGN-SF connection studies with large samples.

This paper is organised as follows. In~\autoref{sec:data}, we describe the S7 survey data and our sample selection.~\autoref{sec:agn_fractions} details the use of BPT diagram mixing sequences to calculate spatially-resolved AGN fractions for \ha{} and \oiii{}. In~\autoref{sec: Stellar age determination}, we extract the stellar kinematics and light-weighted average age of the stellar population under 100 Myr and 1 Gyr using \ppxf{}. In~\autoref{sec:corr}, we calculate the Eddington ratio and SFR for each galaxy in our sample and report on the correlation of these quantities with the stellar population age. In~\autoref{sec:discussion}, we discuss the observed trends and their caveats.~\autoref{sec:conc} presents a summary of our main findings and proposed extensions of our work.     

\section{Data and Sample Selection}
\label{sec:data}

\subsection{The S7 Survey}
\label{subsec:s7_survey}

Our sample consists of 54 galaxies from Data Release 2 of the Siding Spring Southern Seyfert Spectroscopic Snapshot Survey \citep[S7 DR2\footnote{Available at \url{https://miocene.anu.edu.au/S7/Data_release_2/}};][]{dopita2015, thomas2017}. S7 DR2 contains IFU data for 131 active galaxies. The S7 galaxies were selected from the AGN catalogues of \citet{veron2006, veron2010} to investigate the impact of radio jets on the structure and kinematics of the Extended Narrow Line Region (ENLR). S7 galaxies are nearby ($z < 0.02$, $D\lesssim$ 80 Mpc), southern ($\delta < 10^{\circ}$), away from the galactic plane ($|b|\gtrsim 20^{\circ}$) and exhibit radio continuum emission at 20 cm. This selection criterion introduces biases in our sample that limit the applicability of our results to all AGN, especially given that the majority of AGN (>90$\%$) are radio-quiet \citep{padovani2016}. 

The galaxies were observed with the Wide Field Spectrograph \citep[WiFeS,][]{dopita2007, dopita2010} at the ANU 2.3m telescope at the Siding Spring Observatory in Australia. WiFeS has a field of view of $38'' \times 25''$ with $1''$ by $1''$ pixels, which was focused onto the central region of each galaxy. The observations were conducted with the B3000 grating in the blue ($3200 - 5000 \,\si{\angstrom}$, $R = 3000$, $\Delta v \sim 100 \si{km} \si{s}^{-1}$) and the R7000 grating in the red ($5290 - 7060 \,\si{\angstrom}$, $R = 7000$, $\Delta v \sim 40 \si{km} \si{s}^{-1}$). The spatial sampling is finer than 400 pc per \si{arcsec}, resolving the ENLR and sub-kiloparsec features. The median seeing and exposure times were 1.5 \si{arcsec} and 2700 \si{s} for the S7 DR2 sample respectively.

The S7 data was reduced with the PyWiFeS pipeline~\citep{childress2014}, which performs standard image pre-processing such as trimming over-scanned regions and (average) bias subtraction, as well as cosmic ray rejection using a version of LACosmic tailored for WiFeS data. The output of the PyWiFeS pipeline is a calibrated, rectilinear, three-dimensional ($x, y, \uplambda$) grid of data (a “data cube”). 

For the mixing sequence analysis presented in~\autoref{sec:agn_fractions}, we use the emission line fluxes measured in each spaxel from S7 DR2. The emission lines were fitted with 1, 2 and 3 kinematic components using the IDL toolkit \lzifu{}~\citep{lzifu2016}. \lzifu{} recovers accurate emission line fluxes by first fitting and subtracting the stellar continuum using \ppxf{}, and then by carrying out 1-, 2- and 3-component Gaussian profile fits to each emission line. An artificial neural network, \textsc{LZComp} \citep{hampton2017}, was used to determine the number of components required to fit the spectrum of each spaxel in each galaxy. For this analysis, we use the emission line fluxes summed over all kinematic components of each spaxel to maximise the retention of spaxels following S/N cuts. A subset of the 14 Seyfert 1 galaxies in S7 DR2 showed bad emission line fits in the nuclear regions, and we re-fit these regions as described in~\ref{sec:appendix:refitting}.

    For the stellar velocity dispersion and stellar age analyses presented in~\autoref{sec: Stellar age determination}, we used 1\,kpc, 1\re{}, and `S7' aperture spectra provided in DR2, where we first fitted and subtracted emission lines using \lzifu{} prior to using \ppxf{} to analyse the stellar continuum. Further details are provided in Section~\ref{subsec: application_to_s7_galaxies}.

\subsection{Sample Selection}
\label{subsec: Sample Selection}

For each galaxy, we constructed [N\textsc{ii}]/\ha{} vs. [O\textsc{iii}]$\uplambda$5007$\si{\angstrom}$/\hb{} (BPT) and [S\textsc{ii}]/\ha{} vs. [O\textsc{iii}]$\uplambda$5007$\si{\angstrom}$/\hb{} ODDs using spaxels with S/N > 3 on [N\textsc{ii}]$\uplambda$6584$\si{\angstrom}$, [O\textsc{iii}]$\uplambda$5007$\si{\angstrom}$, \ha{} and \hb{}. We also constructed $\mathrm{W}_{\mathrm{H}\alpha}$ vs. [N\textsc{ii}]/\ha{} (WHAN) diagrams using a S/N > 3 cut on [N\textsc{ii}]$\uplambda$6584$\si{\angstrom}$ and \ha{}. We then divided the 131 S7 DR2 galaxies into four categories based on the distribution of spaxels in the BPT diagram: `clean' ($n = 25$), `ambiguous' ($n = 13$), `AGN-dominated' ($n = 16$) and `unsuitable' ($n = 77$). The first three of these categories form our study sample. Clean galaxies satisfy the following 3 key criteria:

\begin{enumerate}
    \item Complete mixing sequence in \nii{}/\ha{} vs. \oiii{}/\hb{}: A mixing sequence must have sufficient spaxels ($\gtrsim 30$) and span the range below the Ka03 line below which galaxies are dominated by star-formation, and above the Ke01 line above which galaxies have a significant contribution from AGN, shocks or HOLMES \citep{singh2013,belfiore2016}. Imposing this criterion ensured that we found pure spectra representative of \hii{} regions and AGN NLR.
    \item Tight mixing sequence: Large scatter in mixing sequences indicates shock excitation or significant variation in ISM properties at fixed AGN fraction \citep{rich2011,kewley2013}, which increases the uncertainty on the AGN fraction measurements.
    \item If present, LINER-like emission is from low-luminosity AGN (LLAGN): Spaxels dominated by LLAGN, shocks, and HOLMES overlap on the BPT and [S\textsc{ii}]/\ha{} ODDs. AGN-dominated emission is often centrally-concentrated, whereas shocks can occur anywhere in the galaxy \citep{rich2011,ho2014,belfiore2016}. Furthermore, the WHAN diagram \citep{cid2011}, which is based on the bimodal distribution of the \ha{} equivalent width $\mathrm{W}_{\mathrm{H}\alpha}$, can differentiate HOLMES from LLAGN. If a galaxy had a significant number of spaxels in the LINER region of the [S\textsc{ii}]/\ha{} vs. [O\textsc{iii}]$\uplambda$5007$\si{\angstrom}$/\hb{} diagram that were spatially extended and $\mathrm{W}_{\mathrm{H}\alpha} > 3\si{\angstrom}$, then it was assumed to have shocks. Galaxies exhibiting extended LINER emission with $\mathrm{W}_{\mathrm{H}\alpha} < 3\si{\angstrom}$ were interpreted as being dominated by HOLMES. Nuclear LINER emission with $3\si{\angstrom} < \mathrm{W}_{\mathrm{H}\alpha} < 6\si{\angstrom}$ was used an indication of LLAGN.
\end{enumerate}

\begin{figure}
\centering
\includegraphics[width=\linewidth]{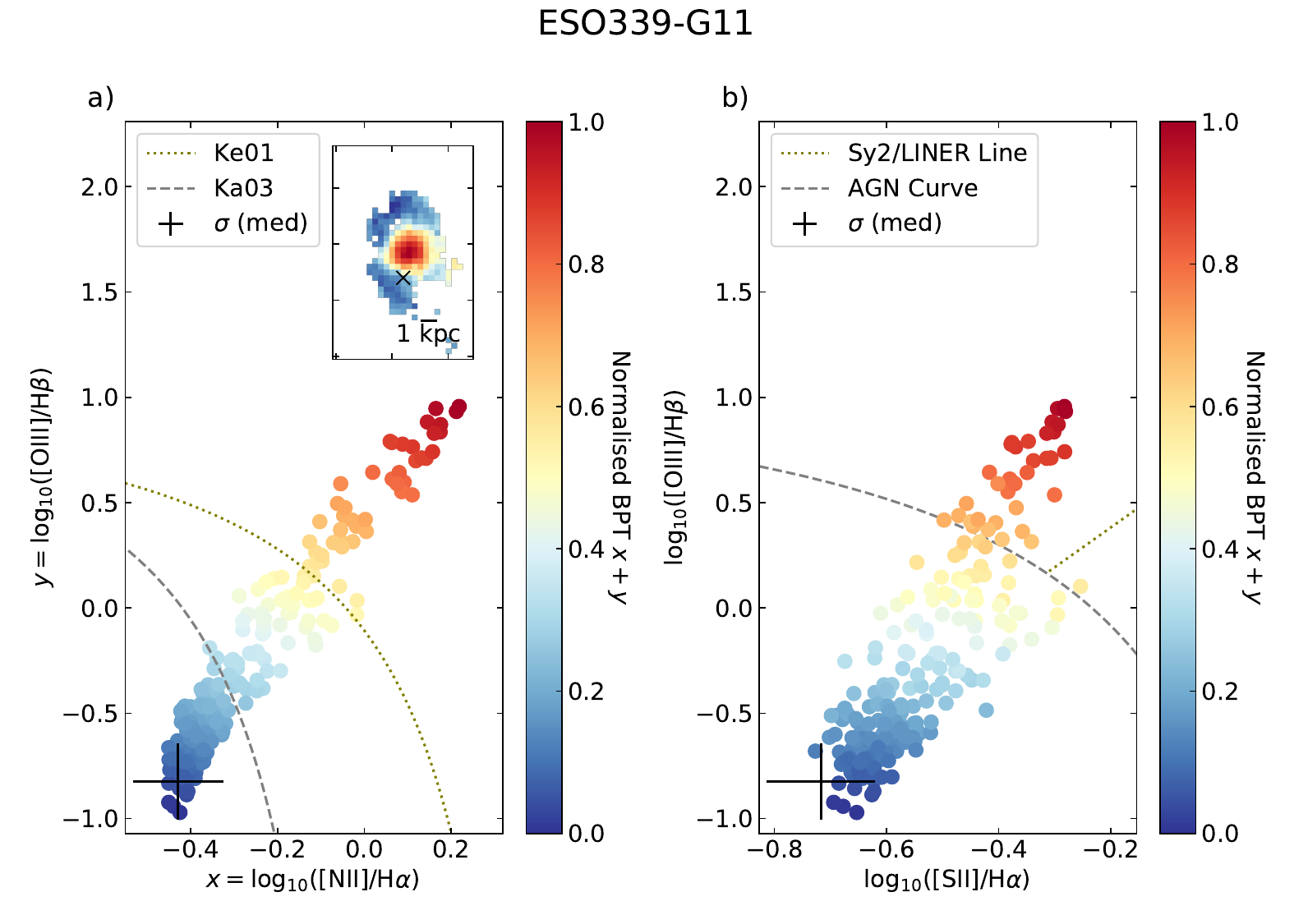}
\caption{BPT diagram (\textit{a}), spatial map (inset in \textit{a}), and [S\textsc{ii}] diagram (\textit{b}) of a representative galaxy from the clean sample. There is a tight and complete mixing sequence extending from the \hii{} region and AGN NLR region. The [S\textsc{ii}] diagram shows data extending to the Seyfert region. The 2D spatial distributions show high line ratios extending radially outwards from the central nucleus, consistent with ionization dominated by the central AGN. Please note that axis limits are chosen individually for each galaxy to avoid compressing the spaxel distributions within a fixed plotting range.
}
\label{fig:clean}
\end{figure}

\autoref{fig:clean} shows galaxy ESO339-G11 from the `clean' sample. Clean galaxies met all three criteria, and contained minimal contamination from shocks or HOLMES. These galaxies showed a visually tight and complete mixing sequence. Additionally, clean galaxies contained a low number of spaxels in the LINER region or the majority of spaxels in the LINER region were associated with LLAGN based on the WHAN diagram and centrally-concentrated LINER emission.

\begin{figure}
    \includegraphics[width=\linewidth]{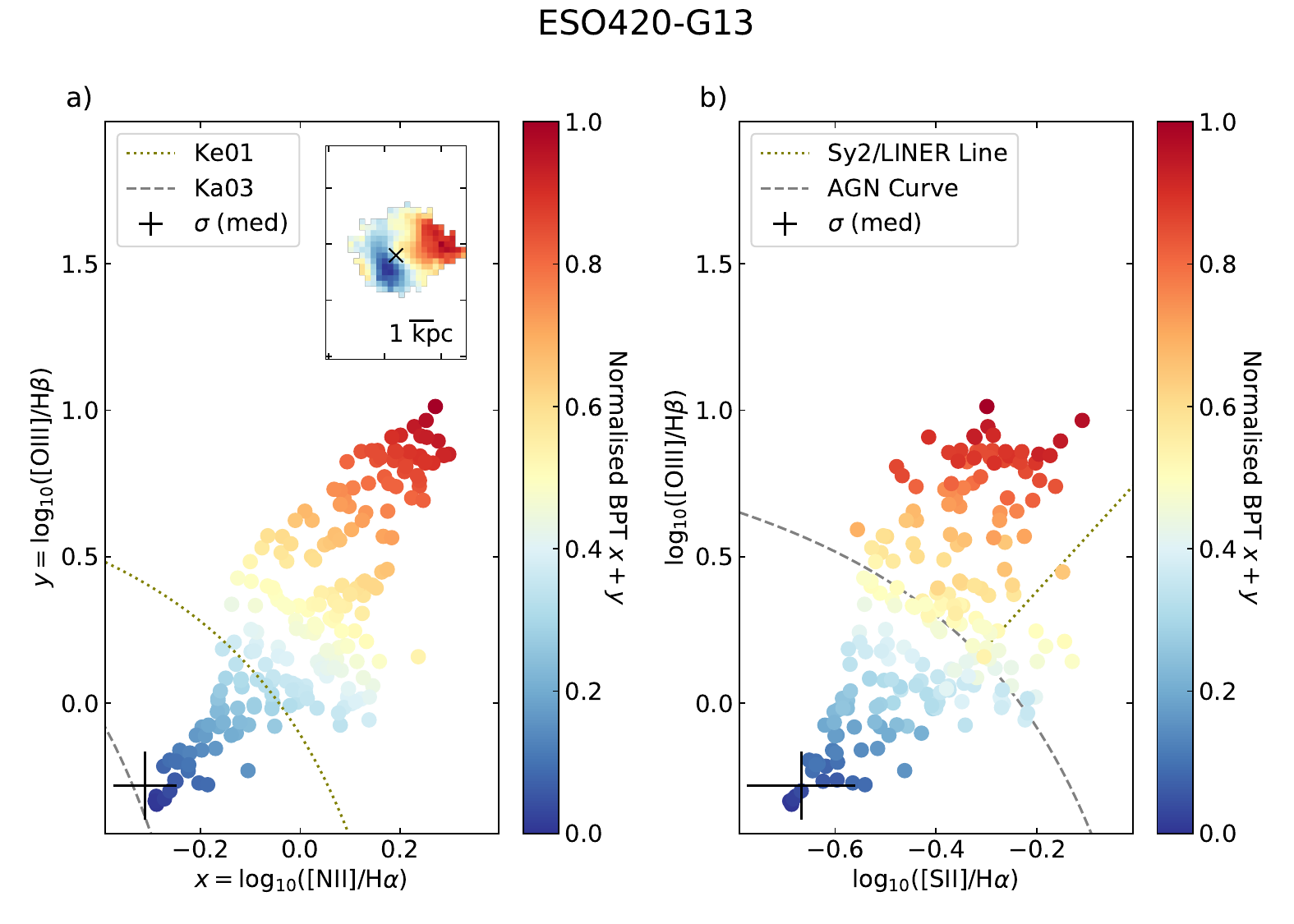}
    \includegraphics[width=\linewidth]{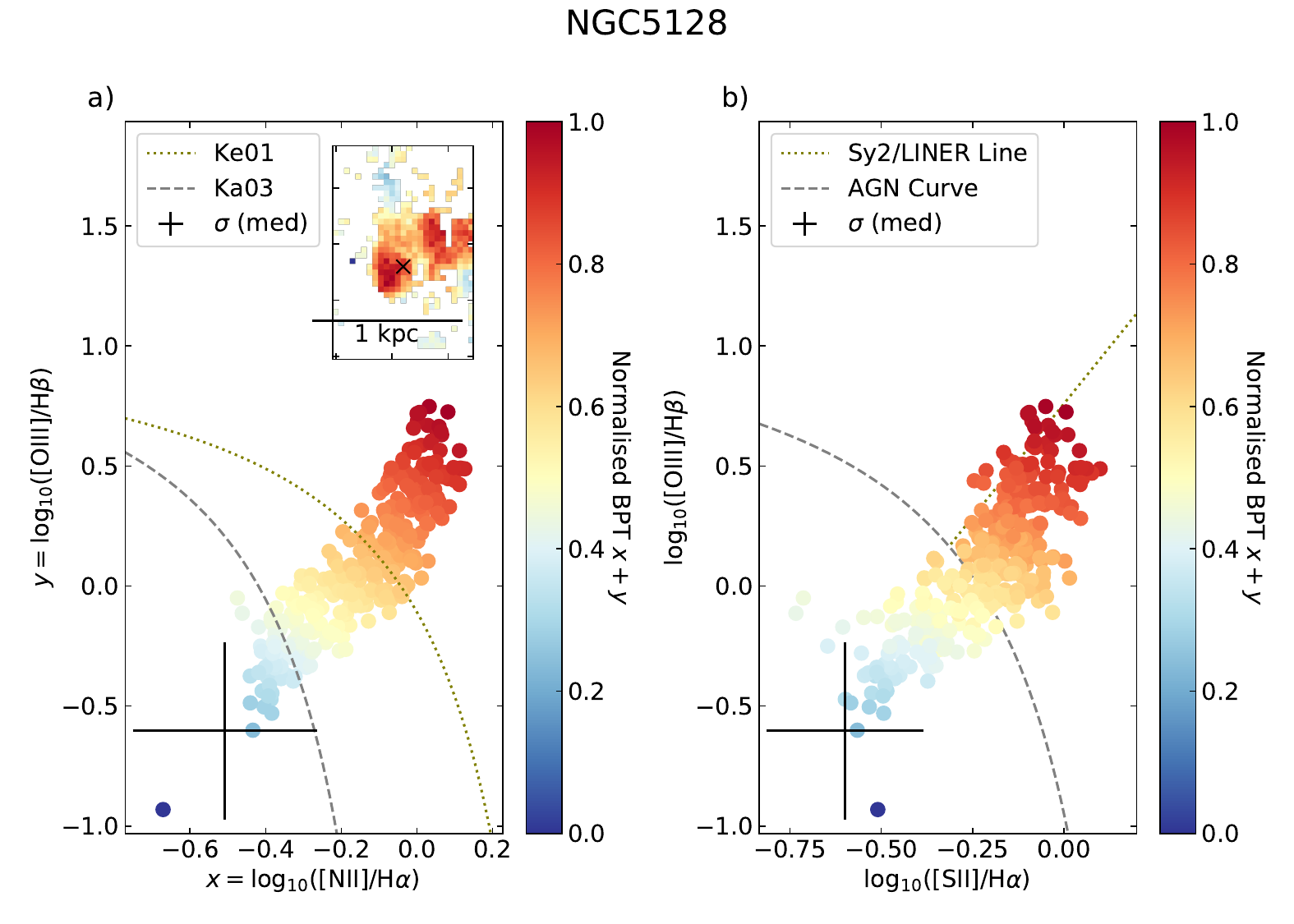}
\caption{Same plots as~\autoref{fig:clean} for two representative galaxies from the ambiguous sample. ESO420-G13 is likely edge-on with the ionization cone pointing to the right. ESO420-G13 may be shock-heated as evidenced by the uneven and dispersed ionization seen in the 2D spatial map and the cluster of spaxels extending off the bottom of the mixing sequence lying in the LINER region. NGC5128 shows a well-defined mixing sequence. However, it contains a high number of spaxels in the LINER region and the same spaxels are widespread throughout the galaxy, consistent with shock excitation.}
\label{fig:ambig}
\end{figure}

\autoref{fig:ambig} shows two representative galaxies, NGC5128 and ESO420-G13, from the `ambiguous' sample. Ambiguous galaxies satisfied the first criterion but may have failed to meet the second or third, suggesting contamination from shocks or HOLMES. We distinguished ambiguous galaxies by the larger spread in their mixing sequence at large line ratios ($\gtrsim 0.25\,\text{dex}$) relative to the clean sample, indicative of shocks (e.g. ESO420-G13,~\autoref{fig:ambig} upper), or a sequence extending towards HOLMES in the WHAN diagram (e.g. NGC5128,~\autoref{fig:ambig} lower). 

\begin{figure}
\centering
\includegraphics[width=\linewidth]{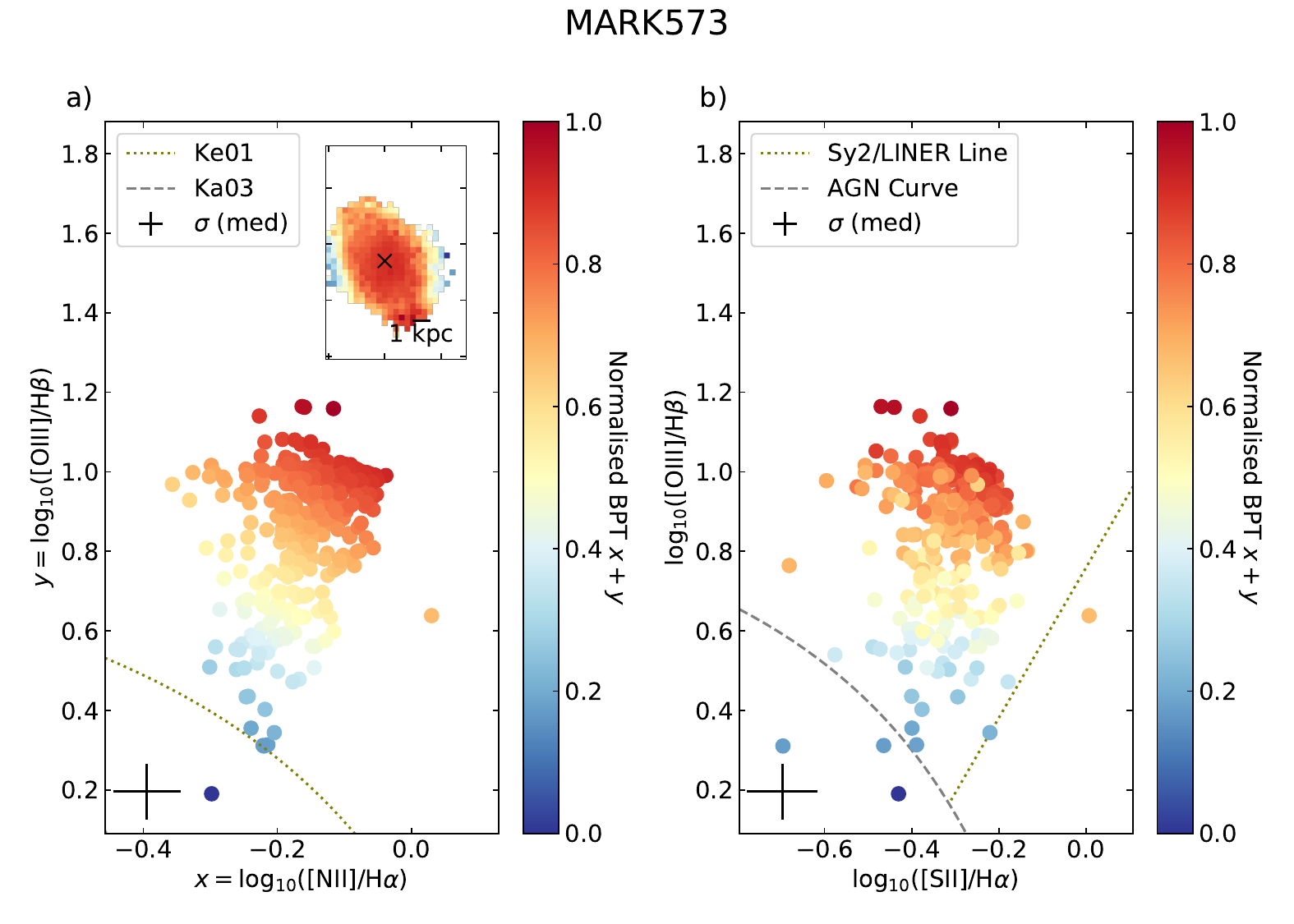}
\caption{Same plots as~\autoref{fig:clean} for a representative galaxy from the AGN-dominated sample. The majority of spaxels lie in the Seyfert region in the [S\textsc{ii}]/\ha{} vs. [O\textsc{iii}]$\uplambda$5007$\si{\angstrom}$/\hb{} diagram.}
\label{fig:agn_dom}
\end{figure}

\autoref{fig:agn_dom} shows galaxy MARK573 from the `AGN-dominated' sample. AGN-dominated galaxies satisfied criteria three and did not need to satisfy criteria one and two. AGN-dominated galaxies did not show optical star-formation signatures. Most spaxels in AGN-dominated galaxies lay above the Ke01 line in the BPT diagram, indicating a significant contribution from AGN or shocks. To separate AGN activity from shocks, we removed galaxies with a high number of spaxels in the LINER region of the [S\textsc{ii}] diagram that were spatially extended. We note that this may have inadvertently removed some LLAGN.

\begin{figure}
    \includegraphics[width=\linewidth]{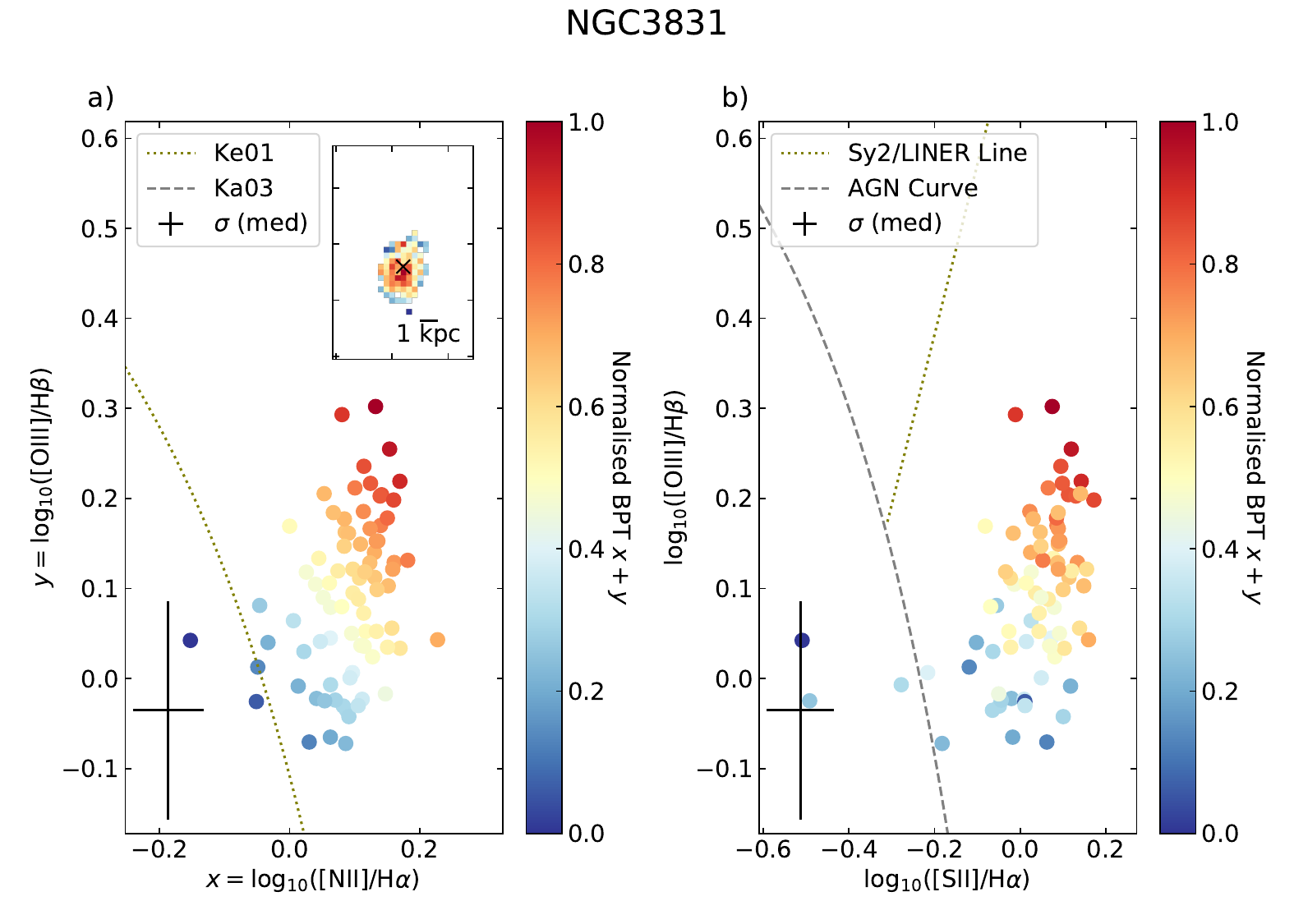} 
    \includegraphics[width=\linewidth]{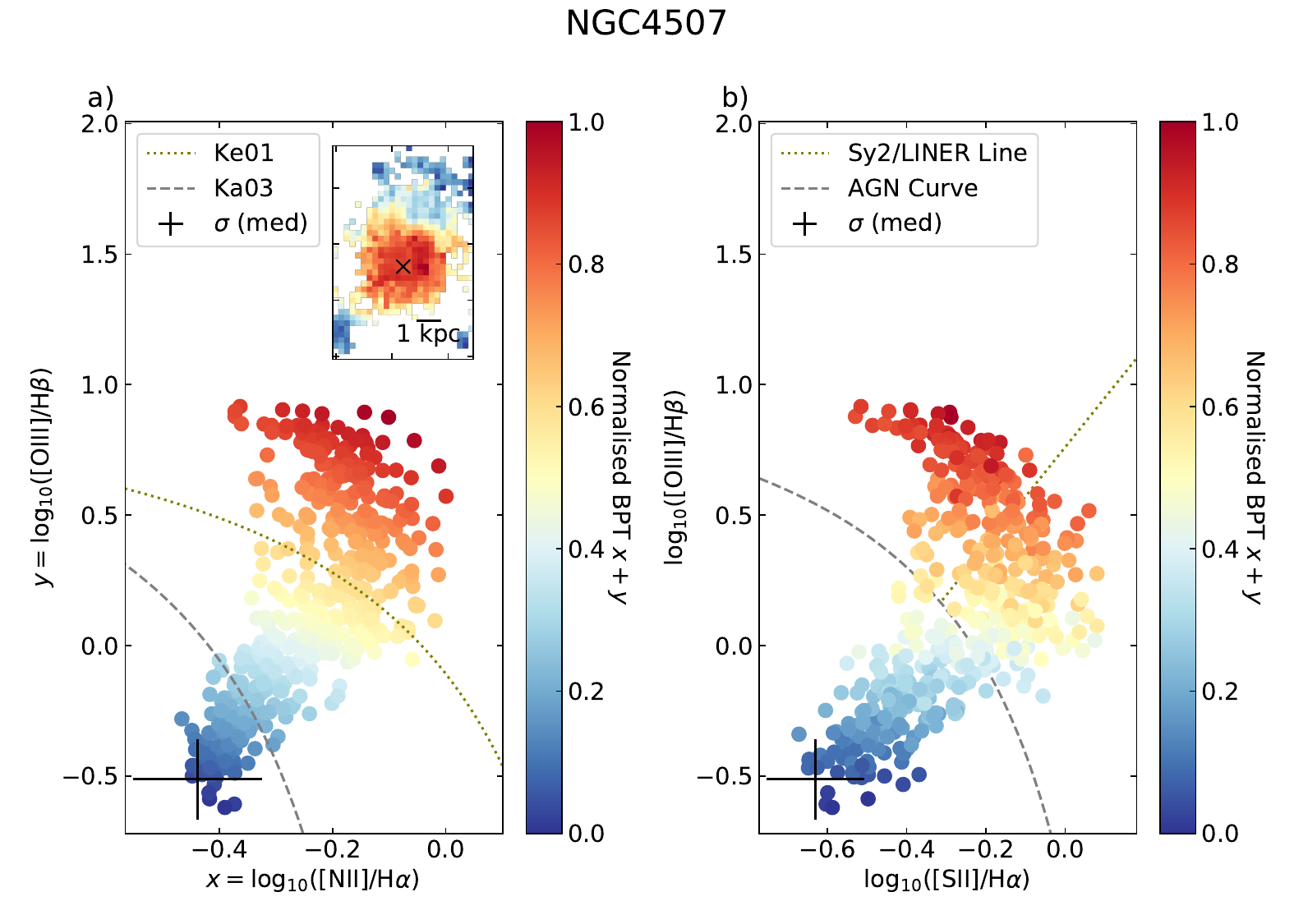}
\caption{Same plots as~\autoref{fig:clean} for two representative galaxies from the unsuitable sample. NGC3831 does not display a complete mixing sequence and is shock-dominated. NGC4507 shows high scatter above the Ke01 line in the BPT diagram, indicating significant shock contamination as the SF+AGN and SF+shock mixing sequences overlap.}
\label{fig:unsuitable}
\end{figure}

\autoref{fig:unsuitable} shows two representative galaxies, NGC3831 and NGC4507, from the `unsuitable' sample. Unsuitable galaxies strongly violated one or more selection criteria, with their spaxels predominantly falling within the star-forming or composite regions of the ODDs, or exhibiting emission primarily driven by shocks or HOLMES (e.g. NGC 3831,~\autoref{fig:unsuitable} upper). Some galaxies were also classed as unsuitable due to their large scatter ($\gtrsim 0.5\,\text{dex}$, e.g. NGC4507,~\autoref{fig:unsuitable} lower).

Our final sample for AGN-SF decomposition consisted of 54 galaxies, categorised into clean, ambiguous, and AGN-dominated classes. Based on optical diagnostic plot classifications derived from the S7 nuclear spectra, \citet{thomas2017} classified 8 of these galaxies as Seyfert 1, 43 as Seyfert 2, 1 as LINER, 12 as starburst (SB) galaxies, 1 as a starburst-only galaxy, and 1 as a post-starburst galaxy. Morphological classifications from HyperLeda indicated that 32 galaxies were spiral, 20 were lenticular, 1 was elliptical, and 1 had an unknown morphology (PKS1306-241).

\subsection{Extinction Correction}
\label{subsec:ec}
Although the ODDs used for our mixing sequence analysis use ratios of closely-spaced lines such as [O\textsc{iii}]/\hb{} and [N\textsc{ii}]/\ha{}, we calculate [O\textsc{iii}] luminosities in Section~\ref{subsec: edd ratio} which rely on an extinction correction. We correct emission line fluxes for interstellar extinction using the Balmer decrement with the MW extinction curve of \citet{fitz1999}. We adopt $(\mathrm{H}\upalpha/\mathrm{H}\upbeta)_{\mathrm{int}} = 2.86$ for the SF-dominated and composite spaxels below the Ke01 line, appropriate for Case B recombination at T = 10,000 K \citep{osterfer2006}, and $(\mathrm{H}\upalpha/\mathrm{H}\upbeta)_{\mathrm{int}} = 3.1$ for the AGN-dominated spaxels above the Ke01 line. Photoionization modelling has shown that shocks, such as those generated by AGN outflows, may increase the Balmer decrement up to $\sim5$ \citep{dop_and_suth_shocks_2017}, introducing systematic uncertainties into the \oiii{} luminosities in some objects.

\section{Measuring AGN Fractions}
\label{sec:agn_fractions}

AGN fractions quantify the relative contribution of AGN to emission line fluxes and probe the strength of the AGN ionizing radiation relative to the stellar ionizing radiation and other significant sources. AGN contributions change with distance from the nucleus and properties of the line-emitting gas such as the metallicity and ionization parameter. We calculate the AGN luminosity of each galaxy in our sample by applying a bolometric correction to the \oiii{} luminosity. In galaxies with strong mixing sequences, a significant fraction of the \oiii{} emission can arise from HII regions. Therefore, we first compute the fraction of \oiii{} emission in each spaxel associated with AGN activity, and then sum the AGN contributions of all spaxels to obtain the total \oiii{} luminosity. Our method described below spatially separates AGN and SF contributions to the \ha{} and \oiii{} lines. Our method involves three steps:

\begin{enumerate}
    \item Removal of outliers in log emission line ratio space using the Mahalanobis distance, which measures the distance of a point from a multidimensional distribution while accounting for correlations between variables.
    \item Calculating SF and AGN basis spectra by averaging values in the lowest and highest of 10 bins, respectively, in $\log_{10}(\textrm{[N\textsc{ii}]/H}\alpha) + \log_{10}(\textrm{[O\textsc{iii}]/H}\beta)$ space.
    \item Constraining superposition coefficients that separate AGN and SF contributions by expressing line luminosities in each spaxel as the linear sum of the basis spectra.
\end{enumerate}

We assume that spectra that lie along a mixing sequence can be decomposed into a linear superposition of basis spectra from a pure \hii{} region and a pure AGN NLR. We use the spectral decomposition method presented in \citet{Davies2016}, where the luminosity of a certain emission line $i$ in a certain composite spectrum $j$, $L_{i,j}$, lying along a mixing sequence can be expressed as
\begin{equation}
    \label{eq: spec decomp}
    L_{i,j} = m_{j, \textrm{AGN}}\times L_{i, \textrm{AGN}} + n_{j, \textrm{SF}}\times L_{i, \textrm{SF}}
\end{equation}
where $L_{i, \textrm{SF}}$ and $L_{i, \textrm{AGN}}$ are the luminosities of the \hii{} region and AGN NLR basis spectra in emission line $i$ respectively. The superposition coefficients $m_j$ and $n_j$ are spaxel-specific weightings that, for a given emission line $i$, quantify the fraction of the luminosity of the basis spectrum that is needed to explain the AGN and SF contribution to the observed luminosity respectively.

We apply this method to the clean and ambiguous classes (described in Section~\ref{subsec: Sample Selection}). For AGN-dominated galaxies, we assign an AGN fraction of 1 and propagate a 30$\%$ flux error since on average galaxies with full mixing sequences intersect the Ke01 line at $f_{\textrm{AGN}} \sim 0.7$. Galaxies with a low number of spaxels ($\lesssim10$) above the Ke01 line are excluded. 

\subsection{Determining Basis Spectra}
\label{sec:bs}

Here we summarise the challenges faced in calculating basis spectra for AGN-SF mixing sequences. We describe our novel method to calculate basis spectra that we have developed in response to these challenges. Our method is empirical, replicable, non-parametric and robust for the wide-ranging mixing sequences in large IFU surveys. 

\subsubsection{Existing Methods to Determine Basis Spectra}

Theoretical photoionization modelling has been used to define basis spectra points of 100 per cent star-formation, shocks, and AGN \citep{dagostino2019}. However, these theoretical models rely on the assumption of parameters associated with the \hii{} region and NLR, such as the temperature and electron density. There is also added error from the inaccuracies in the photoionization models; a 0.1 dex change in the $\log_{10}(\textrm{[N\textsc{ii}]/H}\alpha)$ or $\log_{10}(\textrm{[O\textsc{iii}]/H}\beta)$ ratios of the basis points can change the fractions by up to $\sim$ 10\% \citep{dagostino2019}.

Previous attempts of calculating basis spectra using empirical methods have relied on fitting the mixing sequence in linear emission line space \citep{Davies2016, johnson2023} or simply taking the basis points to be the minimum and maximum [O\textsc{iii}]/\hb{} points \citep{davies2014a, davies2014b}.

The left-hand panel of~\autoref{fig: maha method comparison} shows examples of the fitting method used by \citet{Davies2016} applied to galaxies from our dataset, where the AGN and \hii{} basis spectra are shown in cyan and magenta, respectively. Empirical fitting renders the basis spectra determination to be susceptible to issues such as outliers (e.g. NGC6860, top panel of~\autoref{fig: maha method comparison}), high scatter of the mixing sequence (potentially due to shocks) (e.g. ESO362-G18, second-highest panel of~\autoref{fig: maha method comparison}), the presence of a SF-metallicity sequence due to significant metallicity variations in the SF-dominated regions within a galaxy (e.g.IC1657, second-lowest panel of~\autoref{fig: maha method comparison}) and an uneven distribution of data points (e.g., NGC424, bottom panel of~\autoref{fig: maha method comparison}, which shows a large clustering of points in the pure-SF end of the mixing sequence and sparse points in the pure-AGN end, common when selecting circumnuclear spaxels). 

\subsubsection{A Novel Method of Determining Basis Spectra}

\begin{figure}
    \centering
    \includegraphics[width=0.9\linewidth, height=0.22\textheight, keepaspectratio]{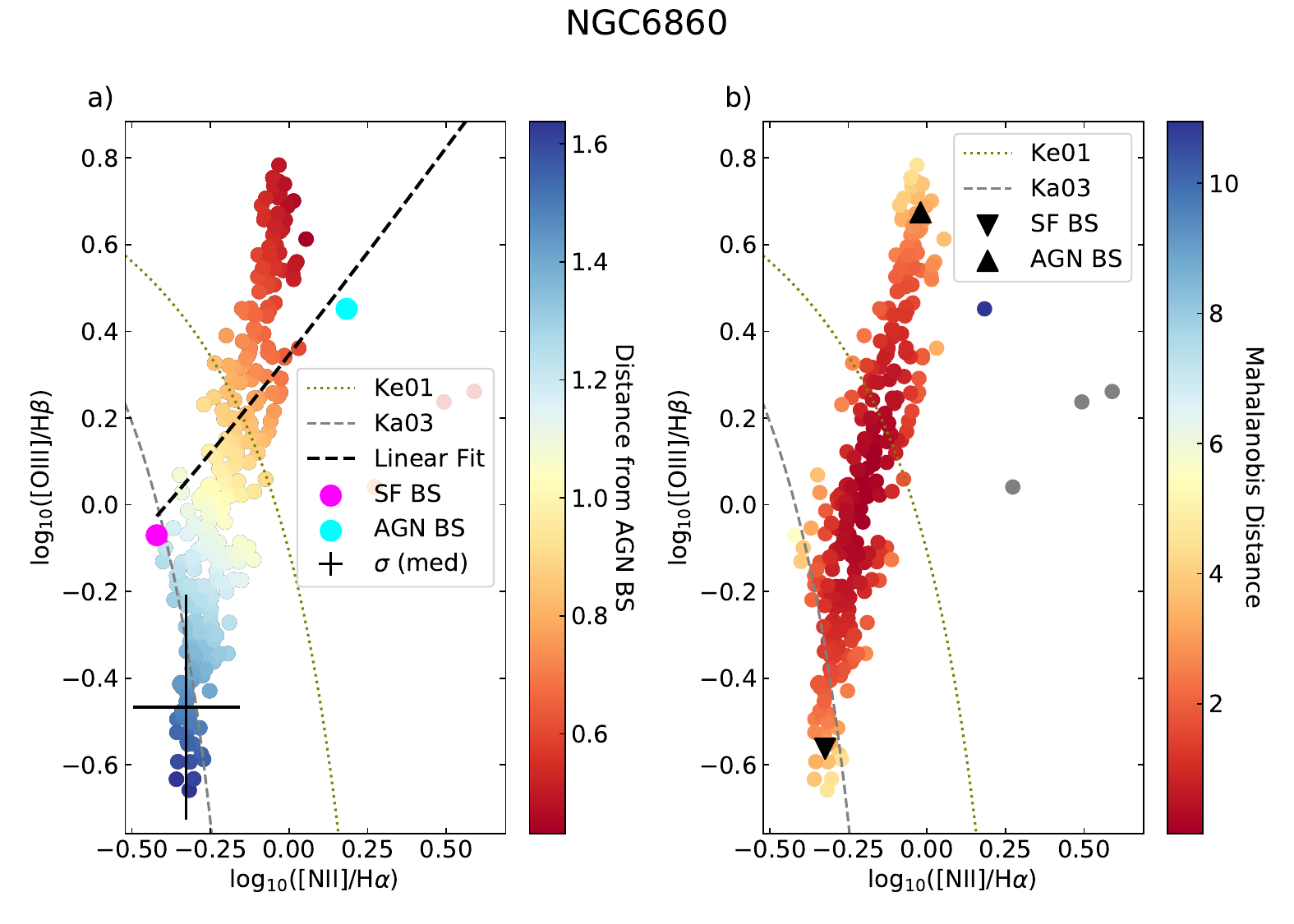} \\
    \includegraphics[width=0.9\linewidth, height=0.22\textheight, keepaspectratio]{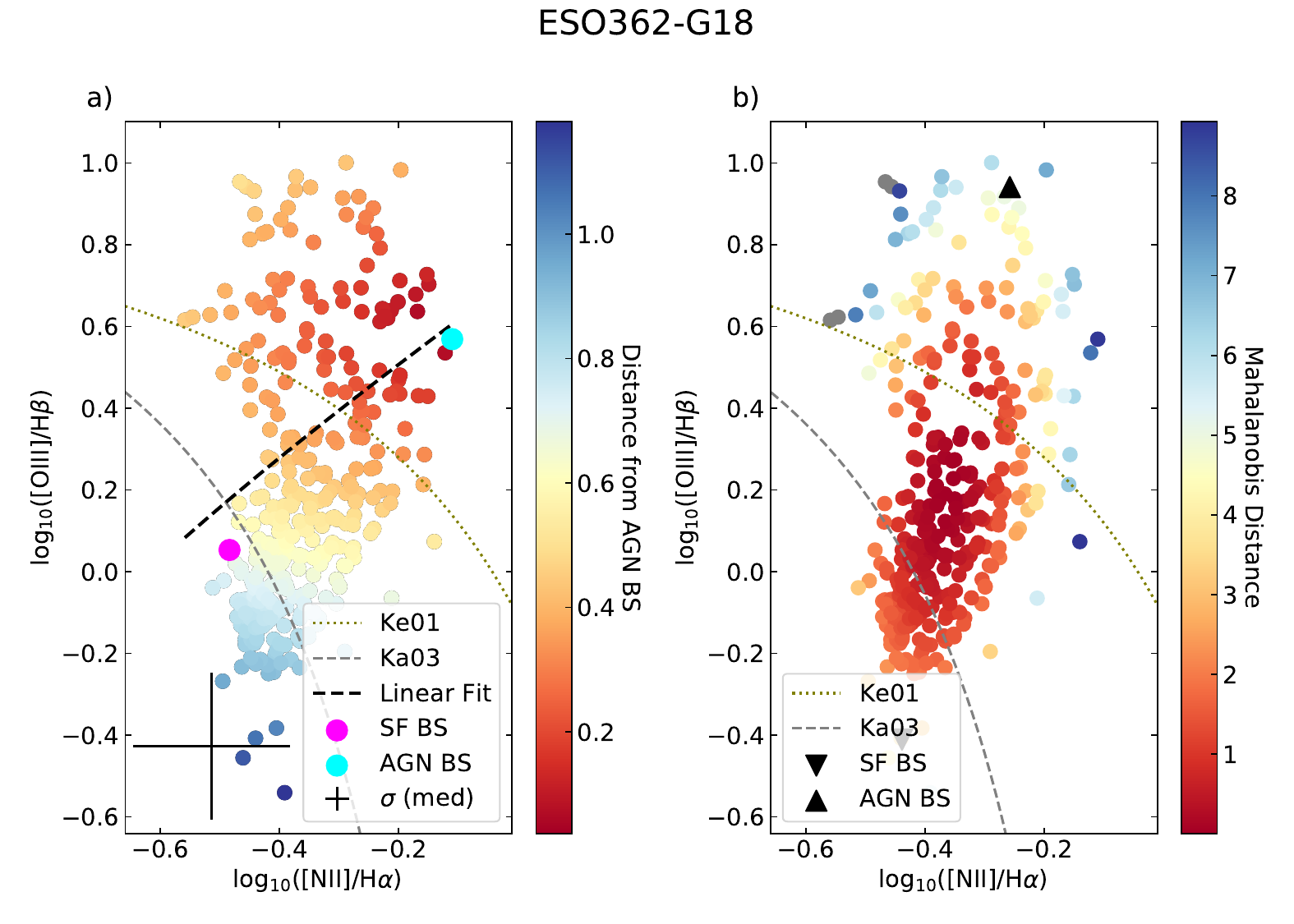} \\
    \includegraphics[width=0.9\linewidth, height=0.22\textheight, keepaspectratio]{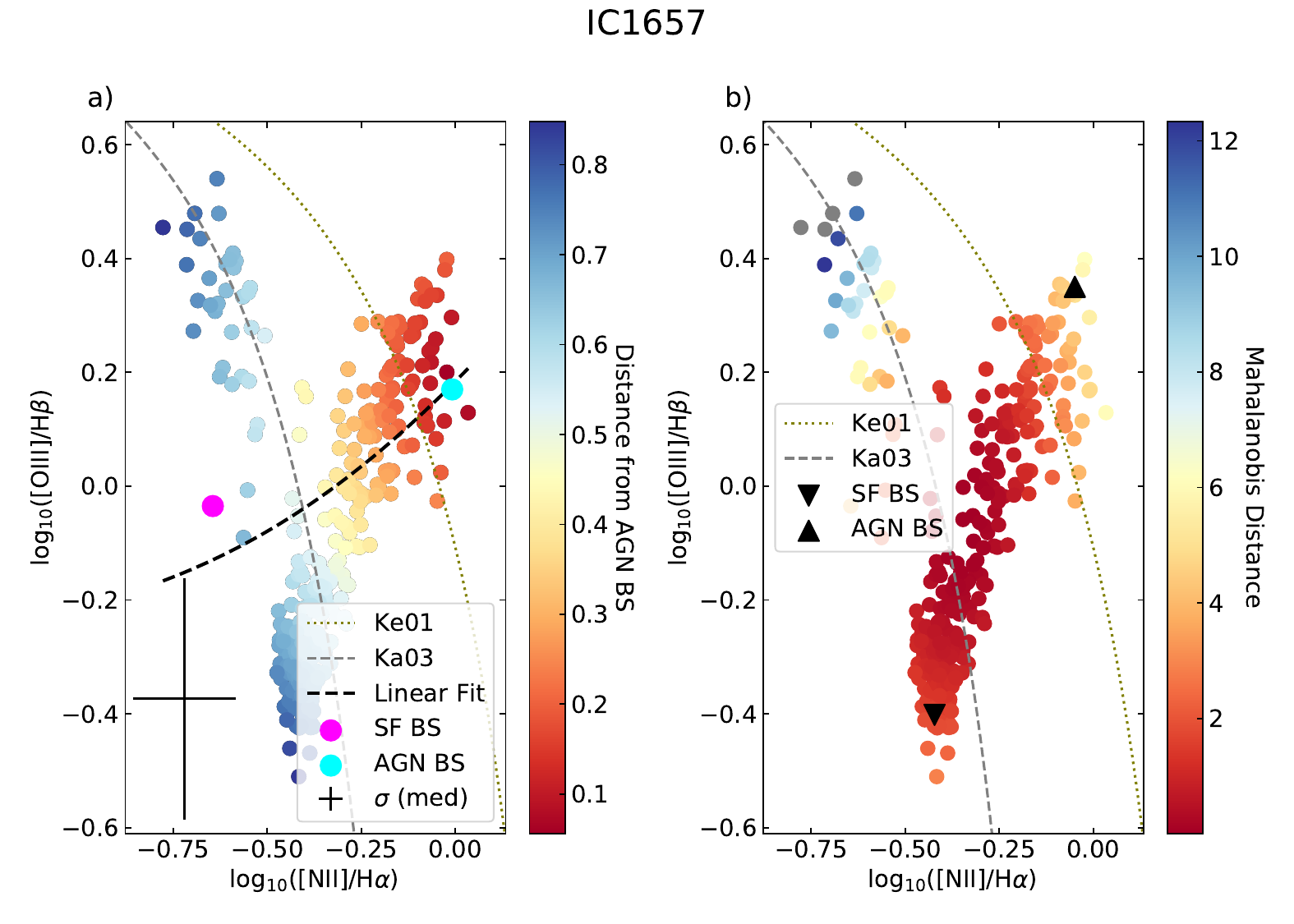} \\
    \includegraphics[width=0.9\linewidth, height=0.22\textheight, keepaspectratio]{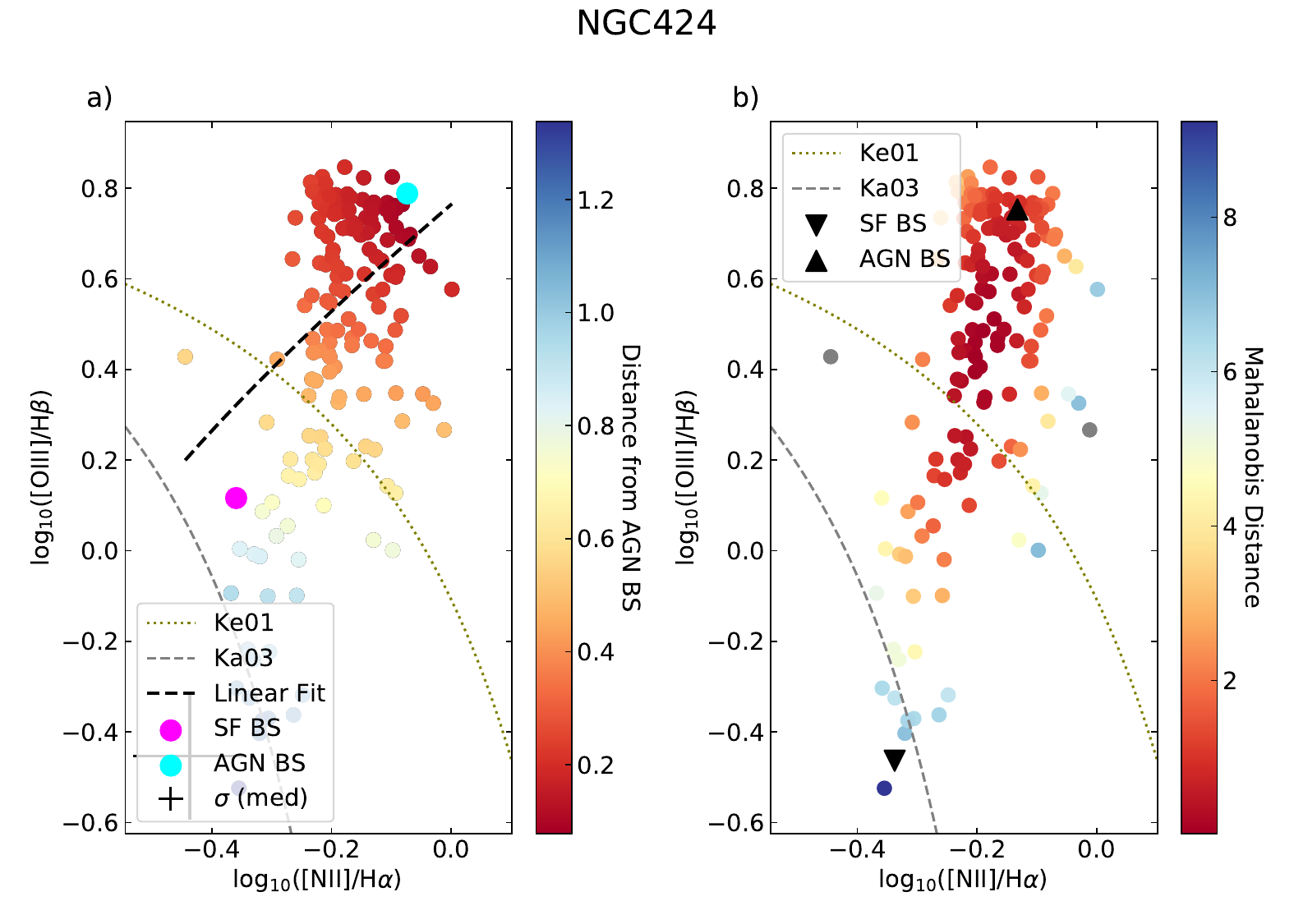}
\caption{A comparison of basis spectra points from the \citet{Davies2016} fitting method (\textit{a}) and the Mahalanobis method (\textit{b}) in each plot pair. The Mahalanobis method is more robust to issues in the linear fit, including outliers (NGC6860), high scatter (ESO362-G18), a strong SF-metallicity sequence (IC1657), and dense clustering in the AGN region (NGC424). Please note that axis limits are chosen individually for each galaxy to avoid compressing the spaxel distributions within a fixed plotting range.}
\label{fig: maha method comparison}
\end{figure}

Here, we present a new method for calculation of the basis spectra that bins data in log emission line ratio space and is much more robust against outliers and variations in the shape and scatter of mixing sequences. The results of applying our new method are shown in the right-hand column of~\autoref{fig: maha method comparison}.

The first step is to remove outliers using the Mahalanobis distance. The Mahalanobis distance is a multivariate distance metric that measures the distance between points and a distribution and is commonly used for outlier detection \citep{maha1999}. The Mahalanobis distance \(D_M\) is defined as:

\[
D_M(\bm{x}) = \sqrt{(\bm{x} - \boldsymbol{\mu})^T \bm{S}^{-1} (\bm{x} - \boldsymbol{\mu})}
\]

where \(\bm{x}\) is the data vector — in our case, the log emission line ratios  
\(\left(\log_{10}(\textnormal{[N\textsc{ii}]/H}\alpha),\ \log_{10}(\textnormal{[O\textsc{iii}]/H}\beta)\right)\);  
\(\boldsymbol{\mu}\) is the sample mean vector and \(\bm{S}\) is the sample covariance matrix.

A key advantage of the Mahalanobis distance over Euclidean distance is that it robustly identifies outliers in the presence of correlations, as is the case in mixing sequences. We calculate the Mahalanobis distance for every point in the BPT diagram. We do not provide errors as weights to compute the covariance matrix \(\bm{S}\) as we assume the position of a data point on the BPT diagram is of higher importance than its uncertainty in defining an `outlier'. To remove outliers, we retain points below the 99th percentile in the Mahalanobis distance, as this cut achieved a good balance in the trade-off between the inclusion of data points outside the mixing sequence (which reduce the accuracy of the basis spectra determination) and the exclusion of end points of the mixing sequence, which are primary candidates for the basis spectra. The points excluded by the Mahalanobis metric are plotted in grey in the right-hand panels of ~\autoref{fig: maha method comparison}.

After removing outliers, the basis spectra are computed as follows. We bin our data using 10 equal-width bins in $\log_{10}(\textrm{[N\textsc{ii}]/H}\alpha)$ + $\log_{10}(\textrm{[O\textsc{iii}]/H}\beta)$ space, using the minimum and maximum values of the data as the lower and upper bounds of the bins. Binning in logarithmic space improves upon existing methods that bin in linear space by producing a more uniform distribution of data along the mixing sequence. We found that $\log_{10}(\textrm{[N\textsc{ii}]/H}\alpha)$ + $\log_{10}(\textrm{[O\textsc{iii}]/H}\beta)$ gives the most visually accurate locations of the basis spectra for a large variety of mixing sequence slopes, while using 10 bins ensures a balance between sampling the distribution well and reducing the scatter in a given bin. The luminosities of the \hii{} region and AGN NLR basis spectra in emission line $i$, $L_{i, \textrm{SF}}$ and $L_{i, \textrm{AGN}}$ in~\autoref{eq: spec decomp}, are then calculated as the average emission line value in the lowest and highest bins, respectively.

The basis spectra using our new method—indicated by black triangles in the right-hand panels of~\autoref{fig: maha method comparison}—are robustly recovered even in cases where the previous linear fitting method performed poorly. It is robust to outliers and is even able to perform well in galaxies where a significant \hii{} region metallicity sequence is observed.

\subsection{Separating SF and AGN Components}
\label{subsec:Separating}

We constrain the superposition coefficients for each spaxel in each galaxy by applying the Levenberg-Marquardt least squares minimisation algorithm using the \textsc{lmfit} package in \textsc{PYTHON} \citep{lmfit}. We fit~\autoref{eq: spec decomp} to the extinction-corrected luminosities of the four strongest emission lines in our data (\ha{}, [N\textsc{ii}], [S\textsc{ii}] and [O\textsc{iii}]). \hb{} is not used because it is redundant with \ha{} due to the extinction correction.  We only consider spaxels in which all 4 lines of the BPT diagram ([N\textsc{ii}], \ha{}, [O\textsc{iii}] and \hb{}) are detected at S/N $ > 3$. We normalise the line luminosities in each spaxel (including the basis spectra) to an \ha{} luminosity of 1. This ensures that \textsc{lmfit} minimises over the relative shape of the emission lines rather than their magnitude and avoids numerical problems associated with very small and large numbers. We convert the observed emission line fluxes to luminosities using $L = 4 \pi D^2F$ where $D \sim cz/H_0$ is the distance to the source from the Hubble–Lemaître law applied at low redshifts as in our sample ($z < 0.02$). We assume a Hubble constant of $H_0 = 70 \,\rm km\,s^{-1}\,Mpc^{-1}$. 

\begin{figure*}
    \centering
    \includegraphics[width=0.95\linewidth]{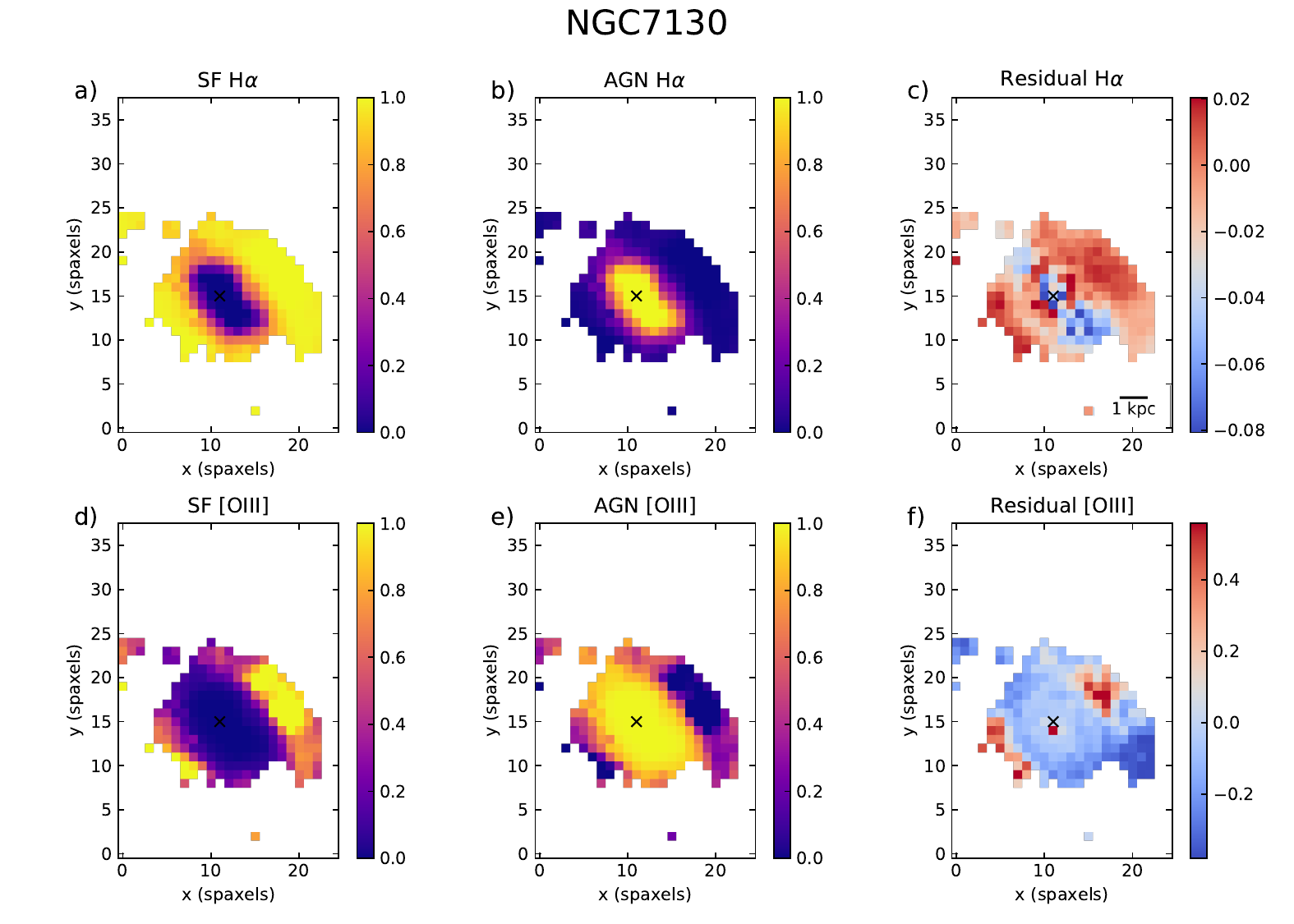}
    \caption{NGC7130 - \textit{a, b}: The resolved SF and AGN fractional contribution to the \ha{} emission line. \textit{c}: The residual (data - model) \ha{} emission. The colour scale has been set to the 1st and 99th percentiles of the residuals to enhance dynamic range. \textit{Bottom}: Same as \textit{a}, \textit{b} and \textit{c} for the [O\textsc{iii}] emission line. All images have been normalised to the total fitted emission and have each spaxel representing an angular size of 1 \si{arcsec} $\times$ 1 \si{arcsec}.} 
    \label{fig:map_agn_frac_ngc7130}
\end{figure*}

We use the superposition coefficients, $m_j$ and $n_j$ in~\autoref{eq: spec decomp}, to construct 2D spatial flux maps that separate the star-forming and AGN components in each emission line. For example, in NGC7130 (\autoref{fig:map_agn_frac_ngc7130}), the H$\alpha$ map traces star-formation in the spiral arms, consistent with HST imaging \citep{malkan1998}, while the AGN H$\alpha$ is confined to the nucleus. The AGN also photoionizes gas in the NLR, producing centrally-concentrated [O\textsc{iii}] emission that dominates the galaxy, with the star-forming [O\textsc{iii}] following the spiral structure.

We provide no spatial information to \textsc{lmfit} in constraining the superposition coefficients and each spectrum is fit independently. Thus, the fact that our decomposition method shows physically consistent spatial distributions of the star-forming and AGN components provides strong evidence that our decomposition method is robust. The bottom panel in~\autoref{fig:map_agn_frac_ngc7130} also includes maps of the residual fluxes (normalised by the fitted flux values), which provide another way of validating the success of the decomposition. The maximum normalised residuals for NGC7130 are on a $\sim 10\%$ level for \ha{} and $\sim 25\%$ level for [O\textsc{iii}]. The results are consistent with the decomposition of NGC7130 from \citet{davies2014a}.

\section{Stellar Continuum Fitting}\label{sec: Stellar age determination}

Star-formation histories (SFHs) and stellar velocity dispersions for the galaxies in our sample were obtained using the \ppxf{} method~\citep{Cappellari&Emsellem2004,Cappellari2017}\footnote{Available: \url{https://www-astro.physics.ox.ac.uk/~cappellari/software/\#ppxf}}, a direct stellar continuum fitting approach, as it leverages the full spectral range of our observations. We use a \ppxf{} fit that is specifically tuned to recover accurate stellar population parameters, rather than use the \ppxf{} fit used by \lzifu{} in Section~\ref{subsec:s7_survey}, which is optimised for accurate extraction of emission line fluxes. To ensure sufficient S/N to enable accurate characterisation of stellar kinematics and stellar populations, we used the 1\,kpc, 1\re and `S7' aperture spectra provided in DR2, where we fitted and subtracted the emission lines using \lzifu{} prior to fitting the continuum using \ppxf{}; further details are provided in Section~\ref{subsec: application_to_s7_galaxies}.

\ppxf{} determines the linear combination of template spectra that best fit the observed spectrum, yielding a non-parametric SFH and chemical enrichment history~\citep[CEH;][]{Cappellari&Emsellem2004,Cappellari2017}. In our implementation, each template represents the spectrum of a simple stellar population (SSP). We used the theoretically-derived SSP templates of \citet{GonzalezDelgado2005} generated using the Padova isochrones, consisting of 74 logarithmically-spaced age intervals from approximately $10^6 - 10^{10}$ yr and 3 metallicities ($0.2, 0.4$ and $0.95\,Z_\odot$), for a total of 222 individual templates. The templates were chosen because at the time of analysis, they were the only templates with sufficiently fine spectral sampling ($0.3$\,\AA\,$\mathrm{pix^{-1}}$) to be reliably convolved to the WiFeS spectral resolution. Since most objects in our sample are massive galaxies that are likely to include an older stellar component, the Padova isochrones were chosen because they include evolution along the red giant branch, unlike the Geneva isochrones. 

We used two different sets of \ppxf\ fits: one optimised to measure accurate stellar velocity dispersions, and a second optimised to obtain accurate SFHs and therefore stellar ages.

\subsection{Stellar Velocity Dispersion Measurements}\label{subsec: Stellar velocity dispersion measurements}

Stellar velocity dispersions were measured following the method of \citet{vandeSande2017}. We used \ppxf\ with a 12th-degree additive polynomial included in the fits to account for calibration errors and stellar template mismatch; no AGN templates or extinction curves were included in the fit, and regularisation was not used. 

First, an initial \ppxf\ to the galaxy spectrum was obtained. In this fit, instead of supplying \ppxf\ with the $1\sigma$ flux uncertainties as the noise, a constant array equal to the average of these uncertainties over the full wavelength range was supplied. The residuals from this first fit were then used to re-scale the $1\sigma$ flux uncertainties; these re-scaled uncertainties were used in all subsequent fits. 

We then ran \ppxf\ 150 times as follows. In each iteration, we took the residuals from the best fit obtained in the initial \ppxf\ run, and randomly shuffled these in wavelength space within 8 equally-sized wavelength bins. These shuffled residuals were then added back to the initial best fit, which was supplied as the input to \ppxf\, as well as the re-scaled flux uncertainties. 

Stellar velocity dispersions ($\sigmastar$) for each galaxy were obtained as the 50th percentile of the distribution of values obtained from the 150 \ppxf\ runs, and uncertainties were obtained by measuring the 16th and 84th percentiles.

\subsection{Stellar Age Measurements}\label{subsec: Stellar continuum fitting with ppxf}

We now describe the \ppxf\ setup that was used to accurately recover the SFH.

Because our targets host AGN, in addition to the stellar templates, we included templates to model an AGN continuum. Following the method of \citet{Cardoso2017}, the AGN continuum was modelled as a power-law using the parameterisation
\begin{equation}
F_\nu(\nu) \propto \nu^{-\alpha_\nu}
\label{eq: AGN power law}
\end{equation}
Four templates were included, with $\alpha_\nu = 0.5, 1.0, 1.5$ and $2$, spanning the range of typical values observed in Seyfert galaxies. Each template is normalised so that $F_\uplambda(\uplambda_{\rm ref}) = 1.0$, where $\uplambda_{\rm ref} = 4020$\,\AA.
The strength of the continuum is given by the sum of the AGN template weights and is parameterised by $x_{\rm AGN}$, given by 
\begin{equation}
x_{\rm AGN} = \frac{F_{\uplambda, \rm AGN}(\uplambda_{\rm ref})}{F_{\uplambda, \rm *}(\uplambda_{\rm ref})}
\label{eq: AGN x_AGN}
\end{equation}
where $F_{\uplambda, \rm AGN}(\uplambda_{\rm ref})$ and $F_{\uplambda, \rm *}(\uplambda_{\rm ref})$ are the template fluxes from the AGN and stellar templates respectively measured at the reference wavelength.

The \texttt{reddening} keyword was used in \ppxf{} to simultaneously fit an extinction curve to the continuum. As is appropriate for star-forming galaxies, we used the \citet{Calzetti2000} extinction curve with $R_V = 4.05$, allowing $A_V$ to vary.
Because an extinction curve is multiplicative, \ppxf{} cannot simultaneously fit a multiplicative polynomial to the solution (which is designed to correct for instrument calibration errors) as the two are degenerate. We therefore did not explicitly account for calibration errors in the fit. 

A Monte-Carlo (MC) method was used to derive the best-fit SFHs for each galaxy. Briefly, \ppxf\ was run without regularisation 1000 times, each time adding random noise to the input spectrum; the final SFH was evaluated as the 50th percentile of the weights in each bin resulting from each \ppxf\ iteration. Uncertainties on the derived weights in each bin were similarly measured using the 16th and 84th percentiles. We also explored using a regularised approach, although we opted to use the SFHs from the MC fits for the reasons discussed in ~\ref{subsubsec: appendix: Regularisation method} and~\ref{subsubsec: appendix: MC simulation method}.

Light-weighted (LW) stellar ages were measured from the best-fit SFHs using 

\begin{equation}
    \log_{10} \tau_{\rm LW}(\tau_{\rm cutoff}) = \frac{\sum_{i}^{i_{\rm cutoff} - 1} w_{i,\rm l} \log_{10} \tau_i}{\sum_{i}^{i_{\rm cutoff} - 1} w_{i,\rm l}},
    \label{eq: LW/MW age equation}
\end{equation}

adapted from eqn. 1 of \citet{McDermid2015}, where $w_{i,\rm l}$ is the weight of template $i$ expressed in $\rm erg\,s^{-1}\,$\AA$^{-1}$ at $4020$\,\AA\ with age $\tau_i$, and we employ cutoff ages $\tau_{\rm cutoff}$ of 100\,Myr and 1\,Gyr to estimate the levels of star-formation occurring on short and long timescales.
We used LW ages rather than than mass-weighted (MW) ages, as the latter are subject to systematic errors caused by uncertainties in stellar mass-to-light ratios, and the former are more sensitive to young stellar populations, which are the focus of this work. 

\subsection{Recovery Tests}\label{subsec: simulations}

There are several systematic effects that can bias stellar age measurements. For our sample, contamination from an AGN continuum is the biggest concern, as our sample comprises both Type\,I and Type\,II AGN with varying degrees of optical continuum emission from the AGN. 
We therefore performed a suite of simulations to quantify how accurately SFHs can be recovered in the presence of an AGN continuum at the spectral resolution and S/N of our data; full details are provided in ~\ref{appendix: ppxf simulations}.

To summarise, we generated mock spectra using SFHs extracted from the cosmological simulations of Taylor \& Kobayashi (in prep.), adding noise to match the approximate S/N of our spectra, as well as varying combinations of extinction and/or AGN continua. SFHs were recovered using \ppxf\ in two different ways: once with regularisation, and a second time using the MC approach described above.

The results are summarised as follows. 
At the S/N of our data, LW/MW ages measured at $\tau_{\rm cutoff} = 100\,\rm Myr$ and 1\,Gyr have approximate systematic uncertainties of 0.2\,dex and 0.1\,dex respectively if no AGN continuum is present.
We also found that \ppxf\ occasionally returns solutions with nonzero young template weights when no young stellar populations are present in the input; we therefore flag age measurements as unreliable if the total light fraction at 4020\,\AA\ in templates below $\tau_{\rm cutoff}$ is below $10^{-3}$. 
Meanwhile, \ppxf\ tends to recover the stellar extinction $A_{V,*}$ to within 0.1\,mag. 

When power-law templates are included in the fit, \ppxf\ accurately estimates the strength of the AGN continuum ($x_{\rm AGN}$, defined as the AGN continuum-to-total light fraction at rest-frame 4020\,\AA). Moreover, if no such continuum is present, \ppxf\ generally reports $x_{\rm AGN} \approx 0$. We therefore assume no optically-significant AGN continuum is present if \ppxf\ reports $x_{\rm AGN} \approx 0$. When there is significant contamination from an AGN continuum, the systematic errors in the LW/MW ages increases to as much as 0.4\,dex depending on the precise shape of the SFH and the slope of the AGN continuum.

We also compared the ages resulting from regularised fits with those from the MC fits; whilst the best-fit SFHs and corresponding age estimates derived using both methods are generally consistent, errors reflecting the effects of random noise are only available when using the MC approach. 
We therefore opted to use the ages derived from the MC fits in our analysis. 

\subsection{Application to S7 Galaxies}
\label{subsec: application_to_s7_galaxies}

Stellar populations are distributed across the galaxy and hence different apertures collect light from different subsets of stars. We calculate stellar ages within three different apertures of varying size: 1 kpc, which probes a constant physical scale typical of bulges \citep{fisherdrory08, gaoetal2020} and circumnuclear star-forming rings \citep{butacrocker1993}; 1 \re{}, the effective half-light radius measured in the blue band of the S7 sample, which accounts for differences in the overall sizes of galaxies with different masses; and the `S7' aperture, a pixel-inscribed circular aperture chosen to approximately match the 3-arcsec-diameter SDSS fibre \citep{thomas2017}. In practice, the S7 aperture spans 4.24 arcsec diagonally and 5 arcsec from the centre spaxel to the end of a row or column.

\begin{figure}
    \includegraphics[width=0.9\linewidth]{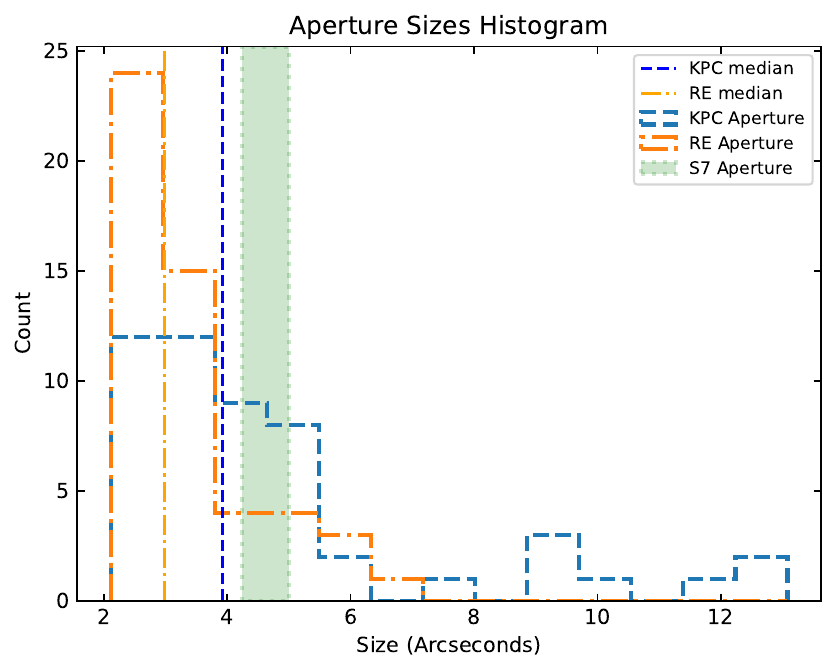} 
\caption{Histogram of the aperture sizes used in our analysis. The S7 aperture ranges from 4.24 arcseconds to 5 arcseconds and is defined in \citet{thomas2017}. Galaxy NGC5128 (Centaurus A) has been removed from this histogram due to its proximity compared to other galaxies in our sample (scale of 1\,kpc $\sim$ 26.45 arcseconds.}
\label{fig:ap_sizes_hist}
\end{figure}

A histogram of the aperture sizes is shown in~\autoref{fig:ap_sizes_hist}. The median aperture diameters are 3.97 and 3.01 and mean aperture diameters are 5.19 and 3.53 arcsec for the KPC and \re{} apertures respectively. As shown in~\autoref{fig:ap_sizes_hist}, some galaxies have very large KPC and \re{} apertures because they are very nearby.

\begin{figure*}
    \centering
    \includegraphics[width=0.7\linewidth]{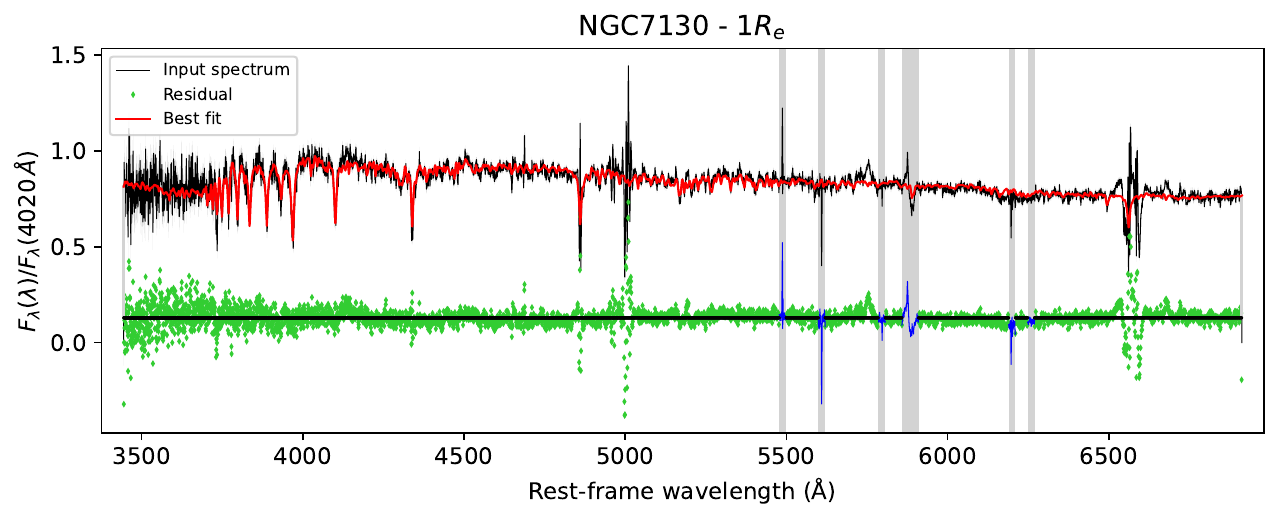}
    \includegraphics[width=0.8\linewidth]{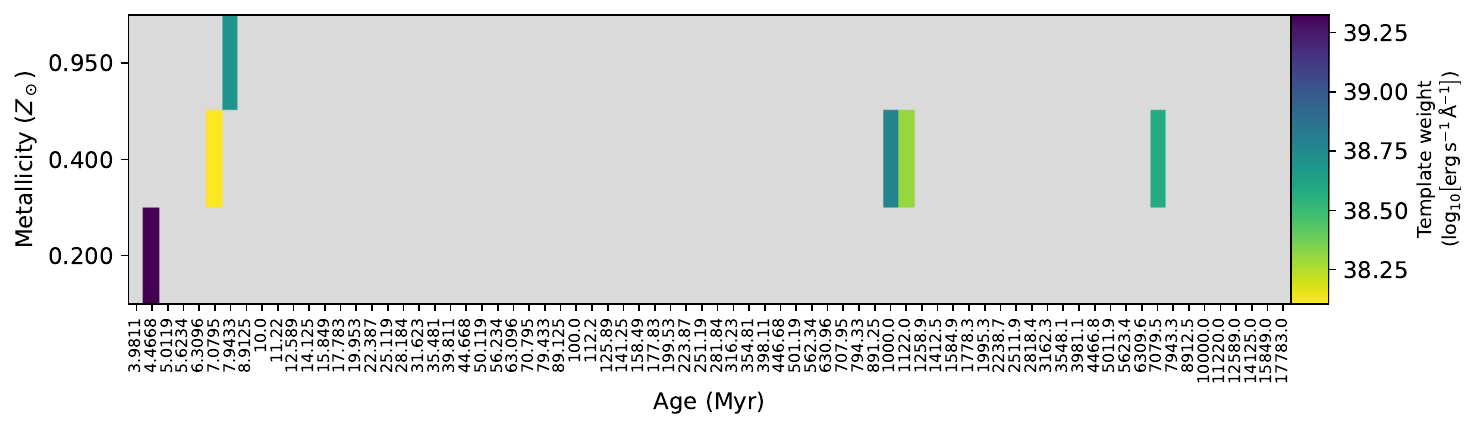}
    \includegraphics[width=0.5\linewidth]{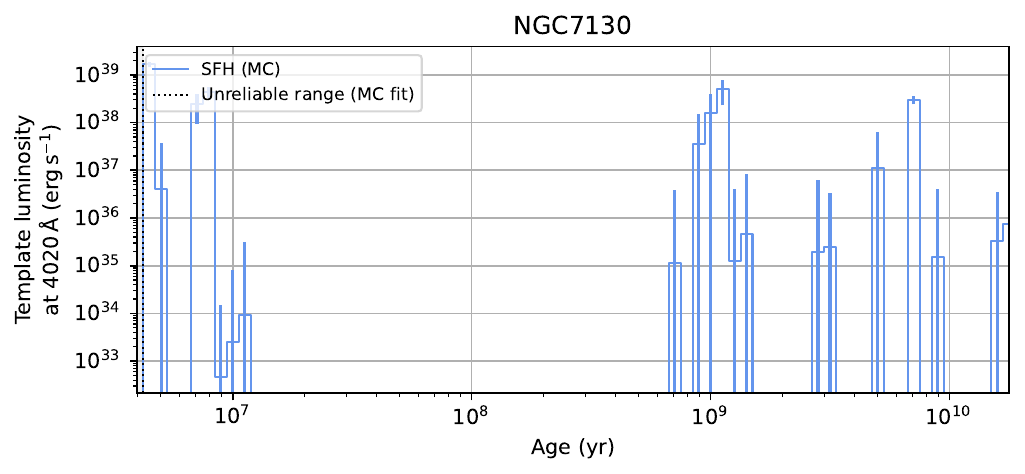}
    \caption{
        Example \ppxf\ fit to the $1\re$ spectrum of NGC7130 taken from one of the MC iterations. The top panel shows the \ppxf\ fit (red) to the spectrum which has had additional noise added (black). The residuals are shown in green, and regions not included in the fit are indicated in grey. The middle panel shows the corresponding best-fit light-weighted SFH and CEH. The bottom panel shows the median SFH from all 1000 MC iterations, where the vertical error bars represent the 68\% confidence intervals on the template weights.
    }
    \label{fig: ppxf example fit - NGC7130}
\end{figure*}

The stellar continuum fits were carried out as follows.
First, continuum-only aperture spectra were created by subtracting emission lines from the multi-component \lzifu\ fits from the 1\re, 1\,kpc and S7 aperture spectra. 
We applied the MC method described in ~\ref{subsubsec: appendix: MC simulation method}, within each iteration adding random noise drawn from a Gaussian distribution with a mean of 0 and a standard deviation equal to the $1\sigma$ uncertainties on the fluxes in each spectral pixel. 
An example fit to the $1\re$ spectrum of NGC7130 and corresponding SFH is shown in~\autoref{fig: ppxf example fit - NGC7130}. Distributions in various parameters resulting from the MC iterations, such as stellar ages, $x_{\rm AGN}$ and $A_V$, are shown in~\autoref{fig: appendix: NGC7130 MC distributions}. 

LW ages were measured for the SFHs from each MC iteration using~\autoref{eq: LW/MW age equation} with $\tau_{\rm cutoff} = 100\,\rm Myr$ and $1\,\rm Gyr$. Final age measurements for each galaxy were taken as the median value across all 1000 iterations, plus associated 16th and 84th percentiles. 
LW ages and the best-fit $x_{\rm AGN}$ and $A_V$ values for each galaxy in our sample are shown in~\autoref{fig: appendix: S7 ppxf galaxy measurements}.

\section{Correlation Analysis}
\label{sec:corr}
In this section, we investigate the relationship between star-formation and AGN accretion by searching for correlations between properties of the stars (e.g. SFR, stellar age) and AGN (Eddington ratio) measured over three apertures: \re{}, KPC and S7.

\subsection{Physical Quantities}
\subsubsection{Stellar Age}
Light-weighted stellar age measurements were obtained for each galaxy using the method described in~\autoref{sec: Stellar age determination} for cutoff ages $\tau_{\rm cutoff} = 100\,\rm Myr$ and $1\,\rm Gyr$ in each of the S7, \re{} and KPC apertures. We added an additional systematic uncertainty of 0.4\,dex to a few light-weighted ages in which the best-fit $x_{\rm AGN} > 0.5$, corresponding to the worst-case effect of a strong AGN continuum from our SFH recovery tests (detailed in~\ref{subsubsec: appendix: The impact of the AGN continuum}). The resulting light-weighted ages are shown in~\autoref{fig: appendix: S7 ppxf galaxy measurements}.

Stellar age measurements were discarded in cases where the total light fraction at 4040\,\AA{} in templates younger than $\tau_{\rm cutoff}$ was below $10^{-3}$; these are indicated in~\autoref{fig: appendix: S7 ppxf galaxy measurements} by the grey points. This affected the following galaxies in at least one aperture and for at least one of the adopted cutoff ages of 100\,Myr and 1\,Gyr: NGC5128, NGC5427, NGC5506, ESO339-G11, IC4777, MCG-03-34-064 and NGC6890.

No age measurements were obtained for IC4329A and NGC7469, both of which exhibited severe emission line residuals which caused the \ppxf\ fits to fail; these galaxies were excluded from our analysis.

\newcommand{\loiiilambda}{\ensuremath{L{\mathrm{[O\textsc{iii}]},\lambda5007}}}

\subsubsection{Eddington Ratio}
\label{subsec: edd ratio}

We use the quantity $\loiiilambda/\sigma_*^4$, hereby simply \los, as a proxy for the Eddington ratio (the accretion rate of the black hole divided by its Eddington accretion rate) as in \citet{ke2006}, chosen because \lothree{} scales with the bolometric luminosity of the AGN for unobscured AGN \citep{heckman04} which scales with the accretion rate. The Eddington rate is proportional to the Eddington luminosity, which is proportional to the mass of black hole, which has been show to scale with the stellar velocity dispersion (the $M_{\textrm{BH}}-\sigma_*$ relation). We adopt $M_{\textrm{BH}} \propto \sigma_*^4$ to stay consistent with \citet{ke2006}, empirical evidence \citep{gebhardt2000} and theoretical arguments \citep{king2003}.

We calculate AGN luminosities by summing the AGN component of the [O\textsc{iii}] emission across the entire WiFeS field-of-view to capture the flux associated with NLR gas and capture extended AGN line emission, which can extend $\sim$20\,kpc or more for ${\mathrm{[O\textsc{iii}]},\lambda5007}$\citep{congiu2017}, for example.

Most of the AGN [O\textsc{iii}] luminosities for our sample lie below $10^{42} \textrm{erg/s}$ as is typical for radio-loud AGN \citep{kukreti2023}. The \los{} values for our sample lie within the range of `composite' and `Seyfert' galaxies compared to figure 20 of \citet{kewley2006} which plots galaxies from SDSS data release 4 (DR4). 

Though limited in size, the AGN-dominated sample shows higher AGN luminosities and comparable median LW ages to the clean and ambiguous categories.

\subsubsection{Star-formation Rate}

The Padova isochrones used in our \ppxf\ analysis in~\autoref{sec: Stellar age determination} are generated from a Salpeter IMF \citep{salpeter1955}. To keep our analysis consistent, we estimate the SFR using the following transformation given by \citet{ken1998}, which is also computed for the Salpeter IMF between 0.1 and 100 $\mathrm{M}_{\odot}$, and for solar metallicity: 

\begin{equation}
    \label{eq:sfr}
  \left(\frac{\textrm{SFR}}{\textrm{M}_{\odot}\,\textrm{yr}^{-1}}\right) = 7.9 \times 10^{-42} \left(\frac{L_{\textrm{H}\upalpha}}{\textrm{erg\,s}^{-1}}\right)
\end{equation}

We take $L_{\textrm{H}\upalpha}$ to be the extinction-corrected SF contribution to the luminosity of the \ha{} line from our decomposition, summed over the \re{}, KPC or S7 aperture introduced in Section~\ref{subsec: application_to_s7_galaxies}.~\autoref{eq:sfr} assumes that no Lyman continuum photons are absorbed by dust \citep{mous2006}, which is transparent at infrared (IR) wavelengths. \citet{kewl2002} showed that \ha{} and IR SFRs agree within $\sim10\%$ if the \ha{} emission line is corrected for extinction using the Balmer decrement and a classical reddening curve, as we do in Section~\ref{subsec:ec}. 

\subsection{Monte Carlo Style Correlation Test}
\label{subsec:mc}

We examine the correlation between star-formation and AGN activity by plotting \los as a function of SFR, age under 100 Myr and age under 1 Gyr measured in the 3 different apertures. We use \lothree{} measured over the entire field-of-view (FOV) and \sigmastar{} that corresponds to the aperture size of the stellar age measurement or SFR.

To quantify the significance of any potential correlations, we calculate the Spearman rank-order correlation coefficient, which is a non-parametric test of how well two variables follow a monotonic function. For each age measurement and aperture, we remove all galaxies for which the uncertainty on the LW age is larger than the computed value. Around 5-10 galaxies or 30-40 per cent of galaxies are removed by this cut on average. 

We use Monte Carlo sampling to test the effect of measurement uncertainties on the observed correlations. We follow the `composite' method outlined in \citet{curran2014} to estimate the uncertainty on the Spearman rank-order correlation coefficient. We briefly describe the method below; the key difference is that we sample without replacement so each data point is used once.

We construct $M = 10,000$ mock datasets, each with length equal to the original dataset. Each mock dataset contains values $x_i$ and $y_i$ where $x_i = X_j + \delta X_j \times \mathcal{G}$, where $\mathcal{G}$ is a random number drawn from a standard normal distribution, $(X_j, Y_j)$ is a random pair from the original data and $(\delta X_j, \delta Y_j)$ is the error of the pair. We sample without replacement to drop the assumption that our sample is representative of a larger `true' population. This is because the S7 sample selects radio-bright AGN and our method to calculate AGN fractions involves eliminating galaxies without complete and low-scatter mixing sequences, resulting in a biased sample. We assign the data points a rank, $Rx_i$ and $Ry_i$, and calculate the Spearman rank-order correlation coefficient and student's t-value for each of these $M$ new data sets. We report the final $\rho$ and $t$ values and their associated error as the mean and standard deviation of the distribution of the relevant quantity.

\subsection{Correlation Results}
\label{sec:Correlation_results}

\autoref{tab:corr_results} shows the Spearman strength-to-error ratio (mean correlation coefficient divided by its standard error) for the Eddington ratio proxy, \los, against stellar quantities calculated over \re{}, KPC and S7 apertures. All of the correlations are relatively weak, reflecting a combination of large intrinsic scatter and large measurement errors, particularly for the stellar ages. 

\begin{figure}
    \includegraphics[width=0.9\linewidth]{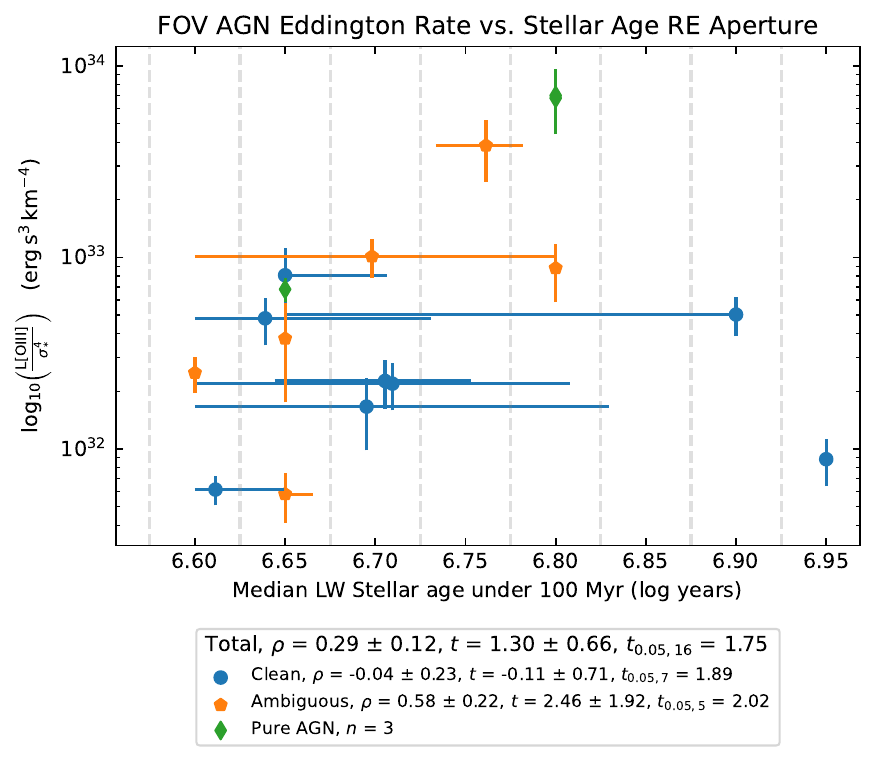} 
\caption{FOV \los vs. the median LW stellar age under 100 Myr, summed over a 1 \re{} aperture. We use \sigmastar{} summed over the same aperture as the stellar age as the differences in $\sigmastar{}^4$ between apertures are on the order of the measurement uncertainty. The vertical grey lines mark the discrete, logarithmically-spaced ages of the SSP templates of \citet{GonzalezDelgado2005}. We also show the t-values and associated critical t-values for a right-tailed test with a 95 per cent confidence interval that quantifies the likelihood of getting a correlation coefficient higher than the observed value by chance. The correlation strengths are reported for the full sample as well as the clean, ambiguous and AGN-dominated subsamples.}
\label{fig:fov_eddvs.re_100Myr}
\end{figure}

\autoref{fig:fov_eddvs.re_100Myr} shows \los{} vs. the median LW stellar age under 100 Myr for the \re{} aperture, which is one of the strongest correlations that we recover.

\begin{figure}
    \includegraphics[width=0.9\linewidth]{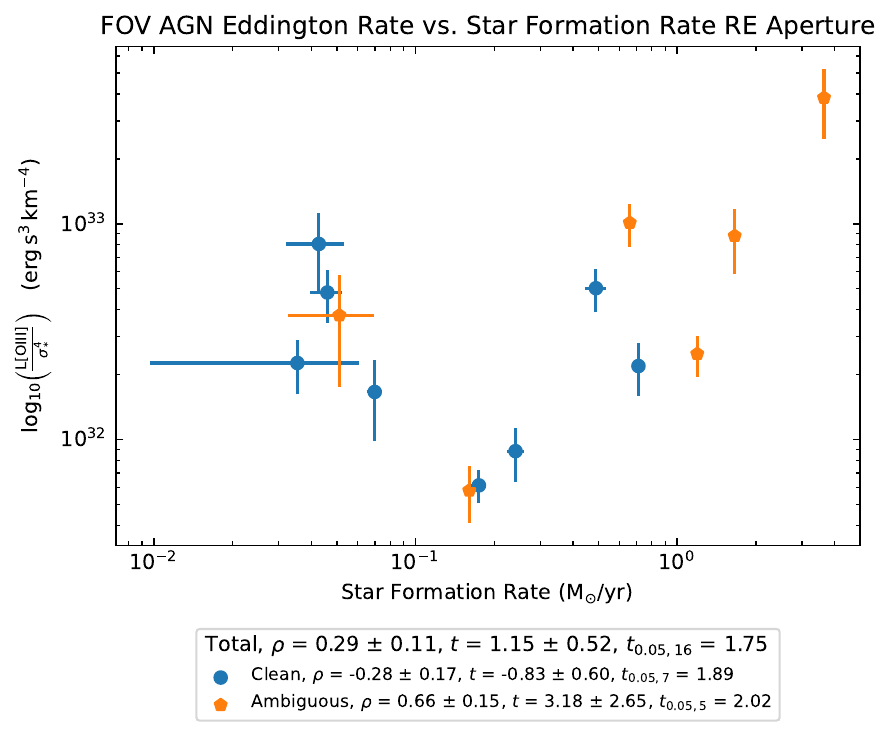} 
\caption{FOV \los vs. SFR, summed over a 1 \re{} aperture. The statistics are reported in the same style as~\autoref{fig:fov_eddvs.re_100Myr}}
\label{fig:fov_eddvs.re_sfr}
\end{figure}

We infer the following results from the data:

\begin{enumerate}
    \item There is a correlation of moderate strength across all apertures between the Eddington ratio and the star-formation rate (see~\autoref{fig:fov_eddvs.re_sfr}). \label{itemone}
    \item For the \re{} and KPC apertures, the Eddington ratio is more positively correlated with the light-weighted age under 100 Myr than the light-weighted age under 1 Gyr. This is a statistically significant effect with an independent two-sample equal variance t-test suggesting that the likelihood of getting different means for the correlation coefficient distributions by chance is $p < 0.01$.    \label{itemtwo}
    \item Weak correlations are observed for the Eddington ratio against the light-weighted age under 100 Myr inside 1 \re{} and the light-weighted age under 1 Gyr inside the S7 aperture. However the latter correlation is primarily driven by an outlier galaxy (NGC5128) with a scale of 1\,kpc $\sim$ 26 arcseconds; following the removal of this outlier, the correlations with stellar age under 100 Myr and 1 Gyr for the S7 aperture are consistent within their uncertainties and thus cannot be considered significantly different. \label{itemthree}
    \item For the SFR and light-weighted age under 100 Myr, the \re{} and KPC apertures show stronger correlations than the S7 aperture.    \label{itemfour}
    \item There are no general trends observed across all apertures when comparing correlations for the clean sample against the combined clean and ambiguous sample. This indicates that including the ambiguous sample does not add any systematic biases to our results. The additional uncertainty on the AGN fractions in the ambiguous galaxies may not have a significant impact on the correlations given the large uncertainty on the parameters derived from the stellar population fitting.  \label{itemfive}
\end{enumerate}

\begin{table}
\centering
\caption{The Spearman strength-to-error ratio (mean correlation coefficient divided by its standard error) obtained using the method outlined in Section~\ref{subsec:mc} for the Eddington ratio proxy, \los, against stellar quantities summed over KPC, \re{} and S7 apertures.}

\begin{tabular}{p{2.75cm}rrr}
\hline
{Aperture} & {\re{} (n = 19)} & {KPC (n = 23)} & {S7 (n = 15)} \\
\hline
LW stellar age <100 Myr & 
{\cellcolor[HTML]{10457E}\color[HTML]{F1F1F1} 3.27} & 
{\cellcolor[HTML]{D5E7F1}\color[HTML]{000000} 1.31} & 
{\cellcolor[HTML]{E58368}\color[HTML]{F1F1F1} -0.47} \\
LW stellar age <1 Gyr & 
{\cellcolor[HTML]{ECF2F5}\color[HTML]{000000} 1.00} & 
{\cellcolor[HTML]{67001F}\color[HTML]{F1F1F1} -1.80} & 
{\cellcolor[HTML]{6EAED2}\color[HTML]{F1F1F1} 2.14} \\
SFR & 
{\cellcolor[HTML]{1F63A8}\color[HTML]{F1F1F1} 3.00} & 
{\cellcolor[HTML]{053061}\color[HTML]{F1F1F1} 3.50} & 
{\cellcolor[HTML]{84BCD9}\color[HTML]{000000} 2.00} \\
\hline
\end{tabular}
\label{tab:corr_results}
\end{table}

\section{Discussion}
\label{sec:discussion}

We find evidence that the AGN Eddington ratio is weakly linked to SFR across all apertures (Result~\ref{itemone}) and young stellar populations for the \re{} and KPC apertures (<100 Myr, Result~\ref{itemtwo}). The 0.2-0.4 dex systematic uncertainties on the LW age under 100 Myr encompass the entire range of values in~\autoref{fig:fov_eddvs.re_100Myr}, making robust analysis challenging. However, the SFRs are much better constrained. Correlations of the AGN Eddington ratio with the SFR and stellar age may be different due to the SFR and stellar age probing SF on different timescales. The \ha-derived star-formation rate is primarily sensitive to young and massive O and early B-type stars, which have lifespans of approximately 10 Myr. In contrast, a galaxy that experienced a burst of star-formation around 50 Myr ago may still exhibit a young LW stellar age, yet show little \ha{} emission, as its O/B stars may have faded.

The trend observed in Result~\ref{itemtwo}, i.e. a stronger correlation between the AGN Eddington ratio and the stellar population under 100 Myr compared to 1 Gyr, is consistent with studies finding higher AGN activity with decreasing stellar age \citet{silverman2009}. The same gas that fuels circumnuclear star-formation may also fuel AGN activity. This timescale is consistent with previous results showing the AGN being fuelled around 50-100 Myr after the onset of star-formation \citep{daviesetal2007} or on the order of 250 Myr after the onset of a starburst \citep{wild2010}.

~\autoref{tab:corr_results} suggests that the correlations in the \re{} and KPC apertures are stronger than the S7 aperture for SFR and LW age under 100 Myr. This may be because the \re{} and KPC apertures are smaller in physical size for most galaxies in our sample (\autoref{fig:ap_sizes_hist}). \citet{delgado2015} measure a larger fraction of young stars and a higher sSFR within 1 \re{} than at larger radii for a diverse set of galaxies from the CALIFA survey. Extending these findings to our sample, Result~\ref{itemfour} may be due to the \re{} and KPC apertures having a younger stellar population compared to the S7 aperture, representing nuclear areas where SF and AGN accretion may be correlated most strongly due to gas inflow towards the nucleus of the galaxy.

Here we explore the viability of stellar mass loss fuelling the AGN. Assuming a SFR of 1 solar mass per year (median SFR of all galaxies summed over the entire FOV), the mass fraction of massive stars between 8 and 100 solar masses being 0.008 under a Salpeter IMF, and 50 per cent of the mass from these massive stars being lost through stellar winds \citep{leith1999}, we estimate a total stellar mass loss of $\sim1.3\times10^{-3} \rm \, M_{\odot}\,yr^{-1}$. We calculate the average BHAR of all galaxies in our sample at all ages to be $\sim5\times10^{-2} \rm \, M_{\odot} \, yr^{-1}$ using the conversion BHAR = $4 \times 10^{-10}$ L[O\textsc{iii}] $\rm M_{\odot} \, yr^{-1}$ from \citet{wild2010} where $L$[O\textsc{iii}] is in units of $\rm L_{\odot}$. Our calculation assumes that the SFR and AGN accretion rate remain constant when in reality we know that both can vary significantly on short timescales. Nevertheless, the stellar mass loss is an order of magnitude too low to explain the observed BHARs alone.

In general the correlations we observe are much weaker than the correlations observed in \citet{wild2010}. This may be because of differences in methods such as our use of IFU data, using radio-selected, actively star-forming galaxies compared to the post-starbust galaxies in \citet{wild2010} and correcting the AGN continuum when calculating stellar indices. We also note that we have a much smaller sample compared to \citet{wild2010}, which leads to less constraining of statistical uncertainties. 

The weakness of the correlations may be attributed to the large observational uncertainties in our measurements, and may also imply that there is not a direct, 1:1 connection between AGN activity and star-formation: a lot of other factors may come into play such as AGN negative feedback suppressing SF. Although the observed trends are very weak, our results are consistent with a picture where stars and AGN grow in lockstep because cold gas is shared between the two processes. 

\section{Conclusions}
\label{sec:conc}

We use a sample of 54 galaxies with IFU data from the S7 survey to investigate the AGN-SF connection in nearby AGN host galaxies. We present a new non-parametric method to compute basis spectra, which we use to decompose the luminosities of individual emission lines at the pixel level into their AGN and SF contributions. We apply this method to compute robust AGN luminosities and SFRs for the galaxies in our sample. We also compute stellar ages under 100 Myr and 1 Gyr by fitting SFHs using \ppxf, accounting for AGN continuum effects, and perform recovery tests to determine the reliability of these measurements. 

We find i) a moderately strong correlation across all apertures between the FOV Eddington ratio and SFR, ii) a significant higher correlation between the Eddington ratio and the stellar light-weighted age under 100 Myr compared to age under 1 Gyr for the \re{} and KPC apertures ($p$ < 0.01), iii) a weak correlation between the FOV Eddington ratio and the light-weighted age under 100 Myr inside 1 \re{} and iv)  a stronger correlation between the FOV Eddington ratio and the SFR and LW age for the \re{} and KPC apertures compared to the S7 aperture. Our results are consistent with previous ideas that AGN are fuelled by recent nuclear star-formation.

The methods applied in this paper can be used to robustly separate AGN-SF interactions for the wide variety of mixing sequences observed in large IFU surveys with AGN such as HECTOR, MaNGA, SAMI, CALIFA and TYPHOON. The upcoming Local Volume Mapper program part of SDSS V offers sub-100 pc resolution for objects out to several Mpc, making it ideal for disentangling AGN and starburst feedback by resolving ionisation structures, shocks, and energy injection scales \citep{lvm2019}. These surveys can complement the S7 galaxies used in our analysis with galaxies that are radio-quiet and span a large range of stellar ages. The Close AGN Reference Survey (CARS) \citep{cars2015, husemanetal2017} is an excellent candidate to extend this study, containing a sample of 41 unobscured (type I) local ($z < 0.06$) AGN. The robustness of our methods against a variety of mixing sequences and SFHs enables application to all galaxies in the sample given certain criteria are met, allowing statistical analyses of the AGN-SF connection with large sample sizes. Applying our empirical approach to large surveys and combining with modelling will allow us to gain better statistics and constrain the physical mechanisms and timescales of the AGN-SF connection.

\begin{acknowledgement}
We thank the referee for their comments, which helped improve the clarity of this manuscript and better situate it within the existing body of work. I would like to thank my mum, Seema Chopra, and my sister, Kashish Chopra, for their endless support; Lisa Kewley for her supervision and guidance during my Honours, of which this work is a continuation; Phil Taylor and Andrew Battisti for discussions on select topics in this paper; Peixin Zhu, who provided \texttt{FITS} files with blank spaxels replaced by a manually chosen n-component fit; and all others who provided feedback on the development of this paper.

RLD is supported by the Australian Research Council through the Discovery Early Career Researcher Award (DECRA) Fellowship DE240100136 funded by the Australian Government.
\end{acknowledgement}


\section*{Data Availability}

All code and plots for this study are available at \url{https://github.com/githubamanc/calc_agn_frac}.

\bibliography{bibliography}

\appendix
\section{Refitting S7 DR2 Emission Lines}
\label{sec:appendix:refitting}

We found the S7 DR2 datacubes to have missing spaxels for 12 out of the 54 galaxies in our subsample. This issue was most prominent at the centre of the galaxies where the AGN dominate the emission line spectra. To obtain emission line fits for these missing spaxels, we refit all 12 galaxies on a spaxel-by-spaxel basis. We use the \lzifu{} setup for the original S7 DR2 run unless specified otherwise. 

Two galaxies, NGC6860 and NGC7469, were identified in the original S7 DR2 paper \citep{thomas2017} as `having very broad and prominent Balmer line components in nuclear spaxels’. These two galaxies were refit on both the regular and broad-component-subtracted cubes. In both cases, unphysical results were produced, with our decomposition emission line maps showing an erroneous SF region between surrounding AGN-dominated spaxels at the centre of the galaxy. Hence these galaxies were excluded from our refitting, but not from our subsample. 
We further excluded 4 galaxies from our refitting sample as the nuclear regions of these galaxies was not fit by \lzifu{} due to a low S/N. We then refit galaxies where there are one or more missing spaxels. 

For the remaining 6 galaxies (IC4995, IC5063, MCG-03-34-064, NGC6890, NGC7130 and NGC7679), we successfully fit the emission lines in the nuclear regions. We found that the orignal S7 DR2 \lzifu{} run was overfitting residuals after the continuum subtraction steps, leading to fitting broad emission lines. This was due to the degree of polynomial used for fitting the continuum-subtraction residuals being too high (\texttt{(b/r)\_resid\_degree} = 15). This parameter was changed so that residuals from the continuum were not fit at all (\texttt{(b/r)\_resid\_degree} = -1). Furthermore, a larger array of initial guesses was provided for the 2nd and 3rd components compared to the original S7 run. In other cases, two or three separate kinematic components were being wrongly fit by a single Gaussian and yielding a local minimum of reduced-chi squared (see figure 3 of \citet{ho2016}). The maximum velocity dispersion constraint for the emission line was increased from 1000km/s to 2500km/s to account for extremely broad lines. Lastly, only those spaxels where the fits failed were fit. This meant that no smoothing refit could be applied (originally applied over 3 iterations and over 3 spatial pixels in the original S7 run) as spaxels were fit independently and may not have had surrounding spaxels in the mask for the refit. The emission line outputs for these galaxies are available upon request and contain two extra extensions: \texttt{MAN\_COMP\_NO}, which contains the visually-selected best number of components to fit the Gaussian flux and \texttt{MAN\_COMP\_USED}, which contains a boolean that when True, indicates that the spaxel has been refit and that \texttt{MAN\_COMP\_NO} should override the best number of components suggested by the artificial neural network in \texttt{COMP\_PREDICTIONS}.

\section{\ppxf\ Recovery Tests}\label{appendix: ppxf simulations}
\input{ppxf_simulations.tex}
\section{Additional \ppxf\ Results}
\input{ppxf_extra_figs.tex}

\end{document}

%% file: ppxf_simulations.tex
\graphicspath{{figs/ppxf/simulations}}
\setlist[enumerate]{labelwidth=0pt}
\setlist[itemize]{labelwidth=0pt}

\subsection{Mock spectra}\label{subsec: appendix: Mock spectra}
\begin{figure*}
    \centering
    \includegraphics[width=0.7\linewidth]{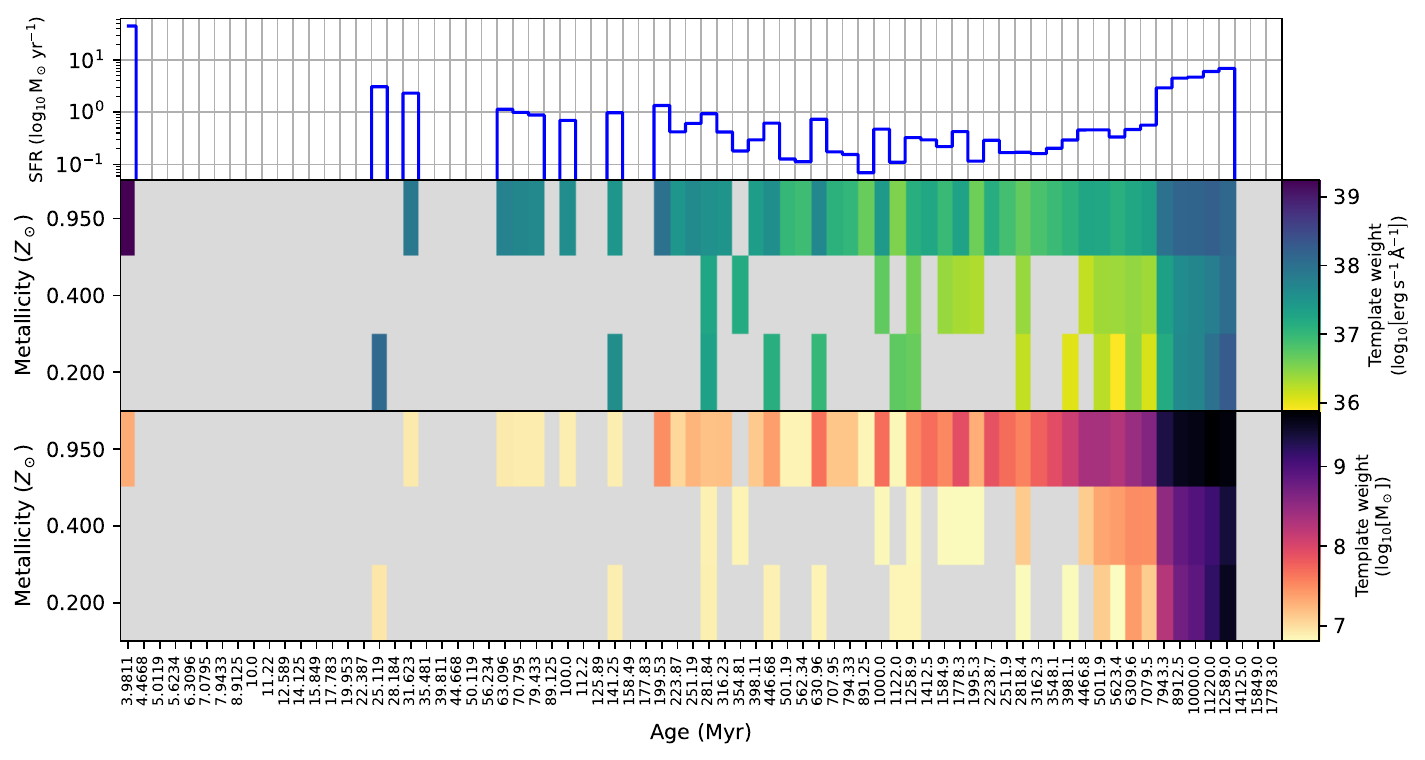}
    \caption{SFH and CEH of ga0044 expressed as template masses (bottom) and template light contributions at 4020\,\AA (middle). The top panel shows the mean star formation rate (SFR) in each template age range. Note that the template weights in ages older than 13\,Gyr are zero due to the age of the cosmological simulation at $z = 0$.}
    \label{fig: ga0044: SFH/CEH}
\end{figure*}

\begin{figure}
    \centering
    \includegraphics[width=0.9\linewidth]{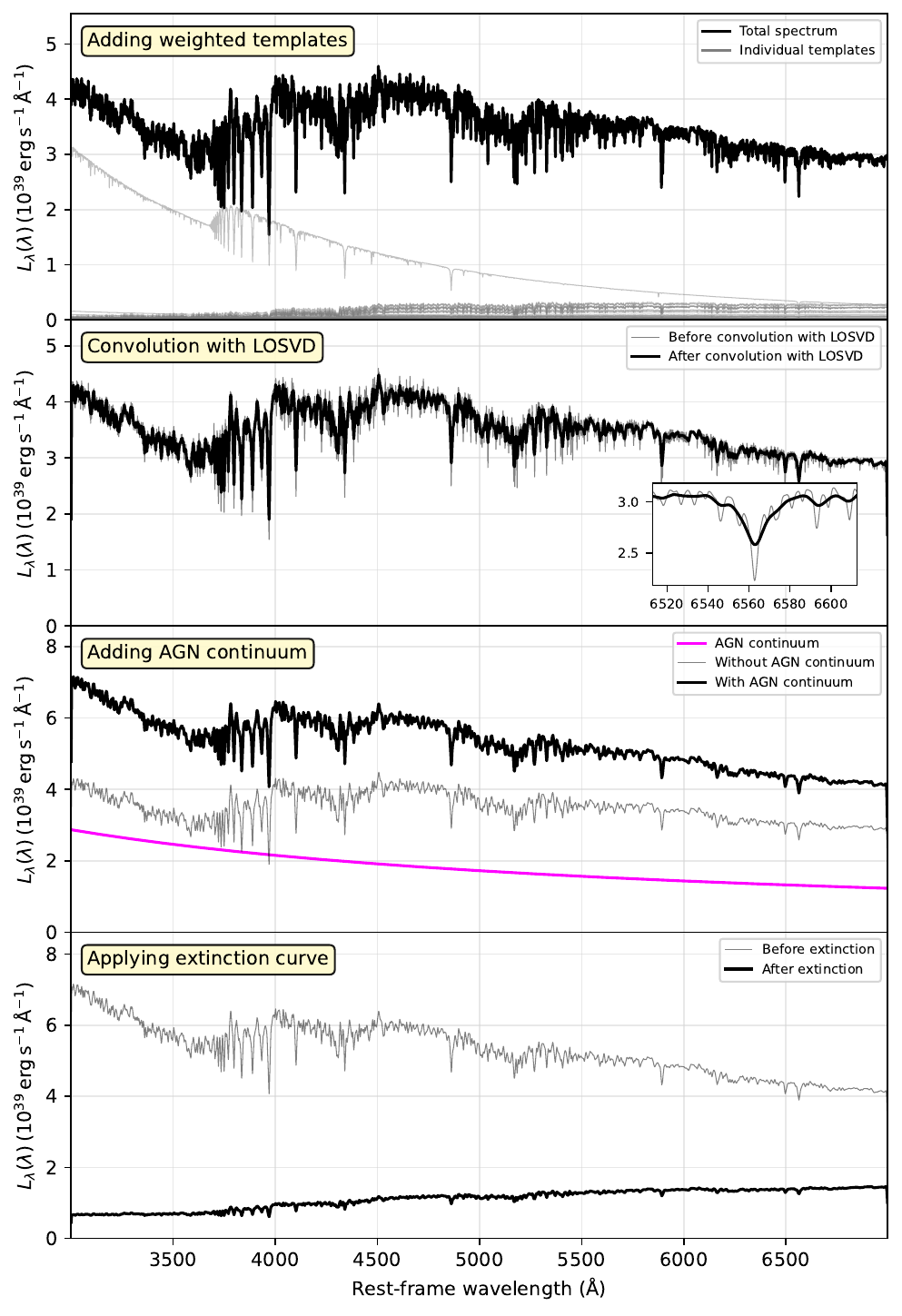}
    \includegraphics[width=0.9\linewidth]{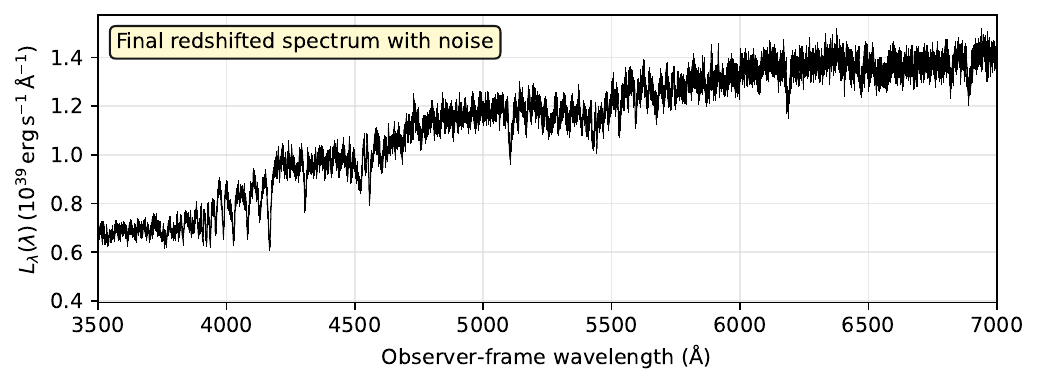}
    \caption{
        Procedure for generating mock spectra as outlined in \ref{subsec: appendix: Mock spectra}.
        First panel, from top to bottom: individual SSP templates, shown in grey, scaled by their weights according to the SFH and CEH of ga0044 shown in ~\autoref{fig: ga0044: SFH/CEH}, and added together to make the mock spectrum (black).
        Second panel: convolving the spectrum with the LOSVD ($\sigma_* = 136\,\kms$). The inset plot shows a detailed view of the wavelength region around the \ha{} absorption feature. 
        Third panel: adding an AGN continuum (pink) with $x_{\rm AGN} = 0.5$ and $\alpha_{\nu} = 1.0$.
        Fourth panel: applying interstellar extinction ($A_V = 1.5\,\rm mag$). 
        Fifth panel: final mock spectrum after redshifting to $z = 0.05$, resampling to the WiFeS wavelength grid and adding Gaussian noise. 
    }
    \label{fig: ga0044: mock spectrum}
\end{figure}

Mock spectra were created using SFHs and chemical enrichment histories (CEHs) taken from galaxies in the cosmological simulations of Taylor \& Kobayashi et al. (\textit{in prep.}) at $z = 0$. Supermassive black hole (SMBH) masses of the simulated galaxies were used to compute realistic stellar velocity dispersions for each galaxy.
Mock spectra were created using the following process:
\begin{enumerate}
    \item The SFHs and CEHs of each galaxy were binned into the 74 age and 3 metallicity bins corresponding to the \citet{GonzalezDelgado2005} templates used in \ppxf{}, generating a $74 \times 3$ grid of mass weights;~\autoref{fig: ga0044: SFH/CEH} shows the SFH/CEH for galaxy ga0044, which has a total stellar mass $M_* = 10^{10.52}\msol$, plus the average star formation rate (SFR) in the age ranges spanned by each template. 
    \item The stellar templates were summed according to these mass weights, yielding the stellar continuum.
    \item The stellar continuum was logarithmically re-binned in wavelength, and then convolved by the line-of-sight velocity dispersion (LOSVD). The LOSVD was approximated as a simple Gaussian, where the velocity dispersion was calculated from the central BH mass using the $M_{\rm BH} - \sigma_*$ relation of \citet{Gultekin2009}.
    \item If desired, an AGN continuum was added. Following the method of \citet{Cardoso2017}, the AGN continuum was modelled as a simple power-law (\autoref{eq: AGN power law}) and the level of the continuum was parameterised by $x_{\rm AGN}$ (\autoref{eq: AGN x_AGN}). 
    \item Extinction was applied to the spectrum using the reddening law of \cite{Calzetti2000}.
    \item The spectrum was redshifted and interpolated to the linear WiFeS wavelength grid using a cubic spline, where the wavelength grid was over-sampled by a factor of 4 to avoid numerical issues during the line spread function (LSF) convolution step. The wavelength grid corresponds to the \texttt{COMB} data cubes provided in DR2 of S7, in which the blue and red data cubes were merged to form a single cube with a spectral resolution corresponding to that of the lower-resolution blue cube ($R \sim 3000$).
    \item The spectrum was convolved by the LSF of WiFeS, which was assumed to be a Gaussian with a FWHM of 1.4\,\AA, as measured from observations of skylines, and down-sampled to the true observer-frame wavelength grid.
    \item Finally, Gaussian noise was added, parameterised by an fixed input S/N ratio at all wavelengths. 
\end{enumerate}
~\autoref{fig: ga0044: mock spectrum} illustrates these steps and shows the final mock spectrum for ga0044. 

\subsection{\ppxf{} fitting method}\label{subsec: appendix: ppxf fitting method}

For our simulations, we used the same \ppxf{} setup as for our observations (see Section~\ref{subsec: Stellar continuum fitting with ppxf}). We explored two different fitting methods: a regularised approach, and a Monte-Carlo (MC) approach.

\subsubsection{Regularisation method}\label{subsubsec: appendix: Regularisation method}
As detailed in \citet{Cappellari2017}, determining the best-fit linear combination of stellar templates is a poorly-conditioned inverse problem, such that there are generally an infinite number of solutions that fit the data to within the provided uncertainties. As such, \ppxf{} can optionally be used with \textit{regularisation}, such that solutions with high-frequency modulations in the SFH are penalised.
The degree of regularisation is controlled by the \texttt{regul} parameter, such that a larger value gives a smoother solution. 
We used the technique recommended by \citet{Cappellari2017} to determine the optimal value of \texttt{regul}, which nominally results in the smoothest SFH that is consistent with the data.



Although this approach ensures the best-fit SFH is the smoothest solution consistent with the data, we found that \ppxf{} struggled to accurately fit SFHs containing both ``spiky'' and ``smooth'' components. Moreover, because the optimal value of \texttt{regul} varies significantly between galaxies and can therefore only be determined via a grid search, it is too computationally expensive to compute uncertainties on the best-fit SFH using this method. We additionally found the regularised fits to be prone to numerical instabilities, in certain cases producing different solutions when run on the same input multiple times.

\subsubsection{Monte Carlo simulation method}\label{subsubsec: appendix: MC simulation method}
To complement our regularised fits, we also used a Monte Carlo (MC) approach to derive best-fit SFHs/CEHs, similar to that used by \citet{McDermid2015}, to compute mass- and light-weighted mean ages for their sample. 
We ran 1000 instances of \ppxf{} without regularisation, each time adding additional noise to the spectrum sampled from a Gaussian distribution with a mean of 0 and a width given by the $1\sigma$ uncertainty on the flux at each wavelength value. 
Because no regularisation is used, the resulting best-fit SFHs/CEHs are much ``spikier'' than those with regularisation; although in most cases the fit appears to be poor, average mass- and light-weighted ages computed from the MC fits tend to have similar accuracies to those computed from the regularised fits. Furthermore, this method enables us to estimate $1\sigma$ errors on the derived quantities by taking the standard deviation in the ages estimated from each individual run.

\begin{figure*}
    \centering
    \includegraphics[width=0.7\linewidth]{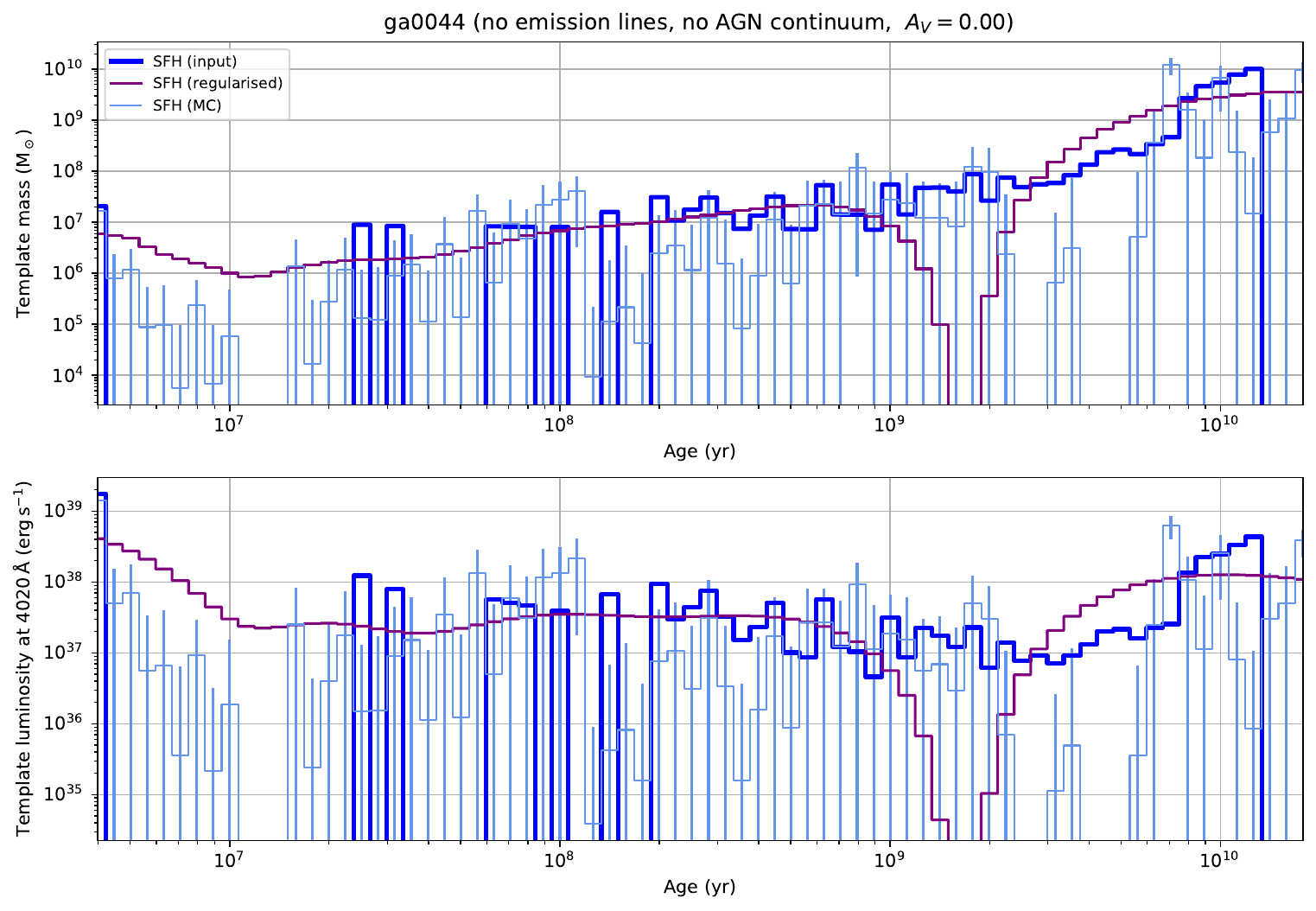}
    \caption{Input SFH of ga0044 (dark blue) and SFHs recovered using \ppxf{} with regularisation (purple) and with MC simulations (light blue), where the error bars represent the standard deviation in the weights from each of the 1000 MC runs. In the top row, the SFHs are expressed in terms of mass per template, and in the bottom row, they are expressed in terms of the template luminosity at $4020\,\text{\AA}$.}    \label{fig:ga0044_input_vs_output_SFH}
\end{figure*}

\autoref{fig:ga0044_input_vs_output_SFH} shows SFHs derived using both regularisation (in purple) and the MC approach (in light blue) for ga0044 in comparison with the true SFH (dark blue) expressed both in terms of template mass and light contribution at $4020\,\text{\AA}$. 
At first glance, neither approach yields a SFH that perfectly traces the input. However, both solutions correctly capture the overall shape of the SFH, and we find that the derived mass- and light-weighted ages are typically accurate to within a few 0.1 dex (See \ref{subsubsec: appendix: The impact of the AGN continuum}).

\subsubsection{Quantifying the age of the stellar population}\label{subsubsec: appendix: Quantifying the age of the stellar population}

Investigating whether there is a link between AGN activity and star formation requires accurate determination of the timescales associated with recent star formation. LW stellar ages ($\tau_{\rm LW}$) were measured using ~\autoref{eq: LW/MW age equation}; we also investigated mass-weighted (MW) ages computed by substituting the mass-weighted template weights $w_{i,\,\rm m}$ for the LW weights in  ~\autoref{eq: LW/MW age equation}.

\begin{figure*}
    \centering
    \includegraphics[width=0.7\linewidth]{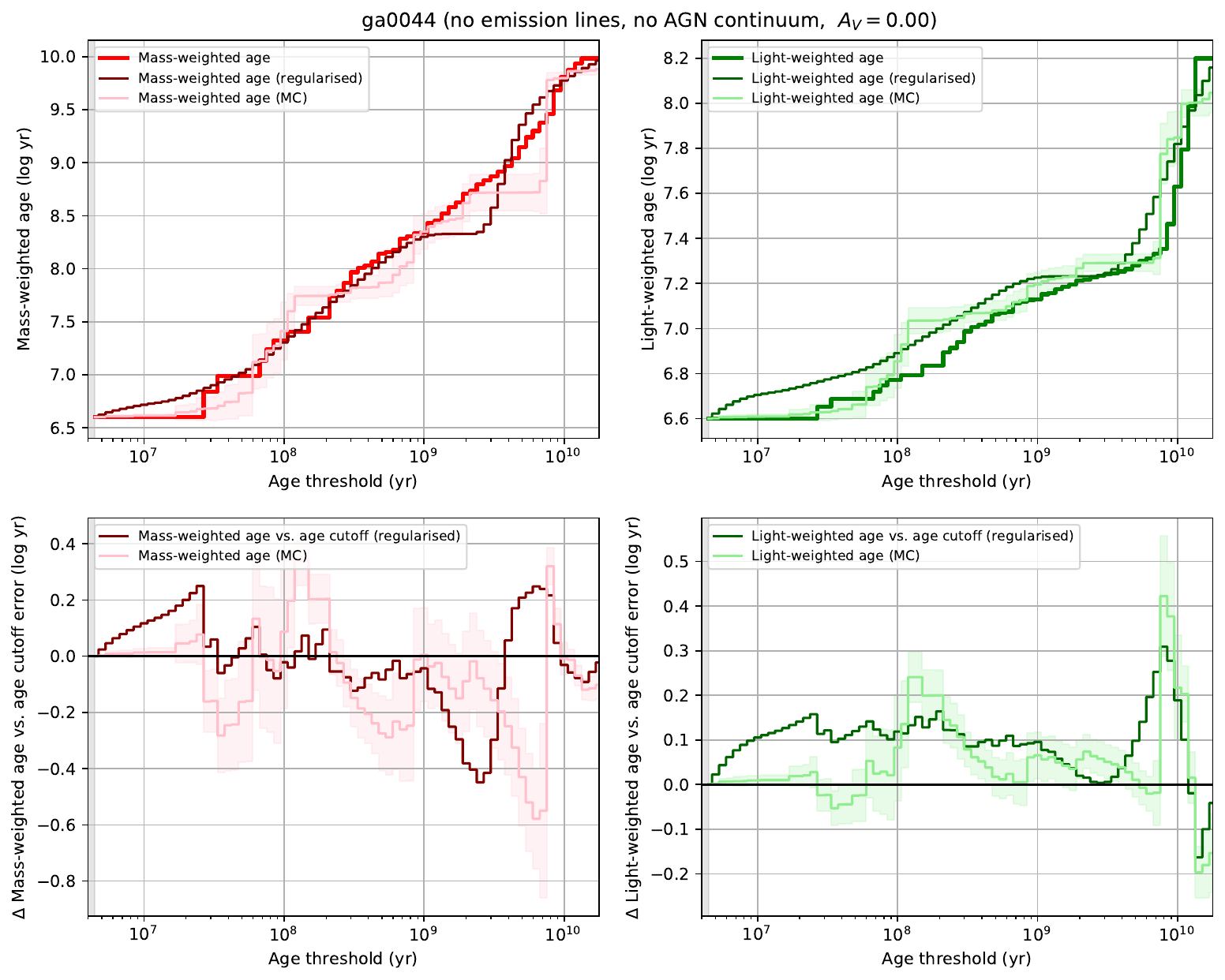}
    \caption{Top row: mass-weighted (left) and light-weighted (right) ages as a function of cutoff age $\tau_{\rm cutoff}$ for the input SFH (thick lines) and the best-fit SFHs from the regularised (thin, dark lines) and MC (thin, pale lines with $1\sigma$ errors shown) \ppxf{} fits for ga0044. The bottom row shows the corresponding errors between the best-fit and input SFHs.}
    \label{fig: ga0044: mass/light-weighted age vs. cutoff age}
\end{figure*}

In ~\autoref{fig: ga0044: mass/light-weighted age vs. cutoff age} we show $\tau_{\rm MW}(\tau_{\rm cutoff})$ and $\tau_{\rm LW}(\tau_{\rm cutoff})$ as a function of cutoff age for the SFH of ga0044, computed from both the regularised and MC fits.
In this particular case, the systematic error in the \ppxf{} fits varies as a function of age; at young ages, \ppxf{} over-predicts both the mass- and light-weighted ages by up to 0.2\,dex, whilst for cutoff ages between $10^9 - 10^{10}\,\rm yr$, the mass-weighted ages are under-estimated by up to 0.6\,dex whereas the light-weighted ages are over-predicted by up to 0.4\,dex. 

\subsection{``Best-case'' scenario}\label{subsec: appendix: "Best case" scenario}
Using the method described in \ref{subsec: appendix: Mock spectra}, we created mock spectra for 106 of the most massive simulated galaxies, spanning a mass range similar to that of the S7 sample. To simulate the ``best-case'' scenario, we added no AGN continuum, and assumed a constant S/N of 100, which is typical of the S7 sample. 
We ran \ppxf{} on each spectrum using both the MC and regularised approaches described in \ref{subsec: appendix: ppxf fitting method}, and computed the mass- and light-weighted ages at 100\,Myr and at 1\,Gyr using the method described in \ref{subsubsec: appendix: Quantifying the age of the stellar population}.

\begin{figure*}
    \centering
    \includegraphics[width=0.4\linewidth]{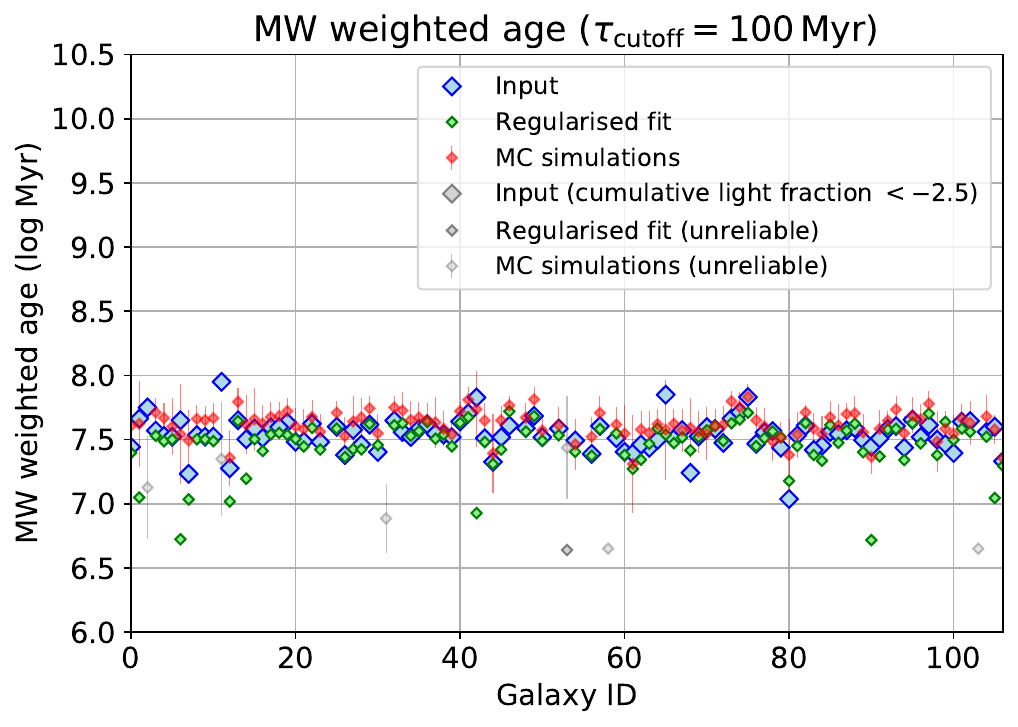}
    \includegraphics[width=0.4\linewidth]{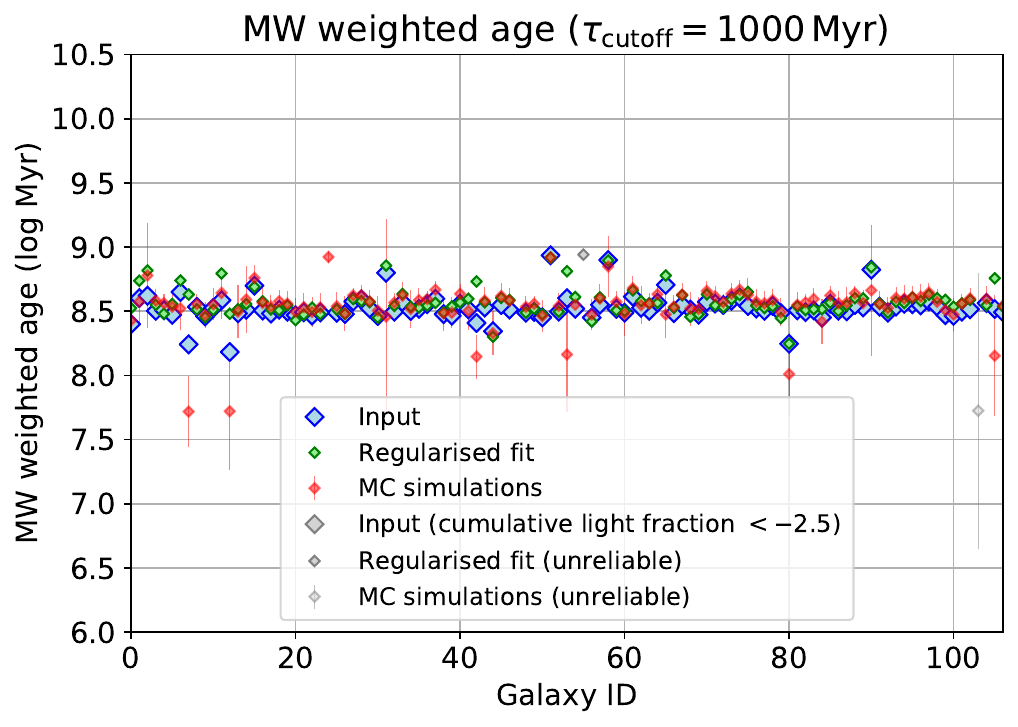}
    \includegraphics[width=0.4\linewidth]{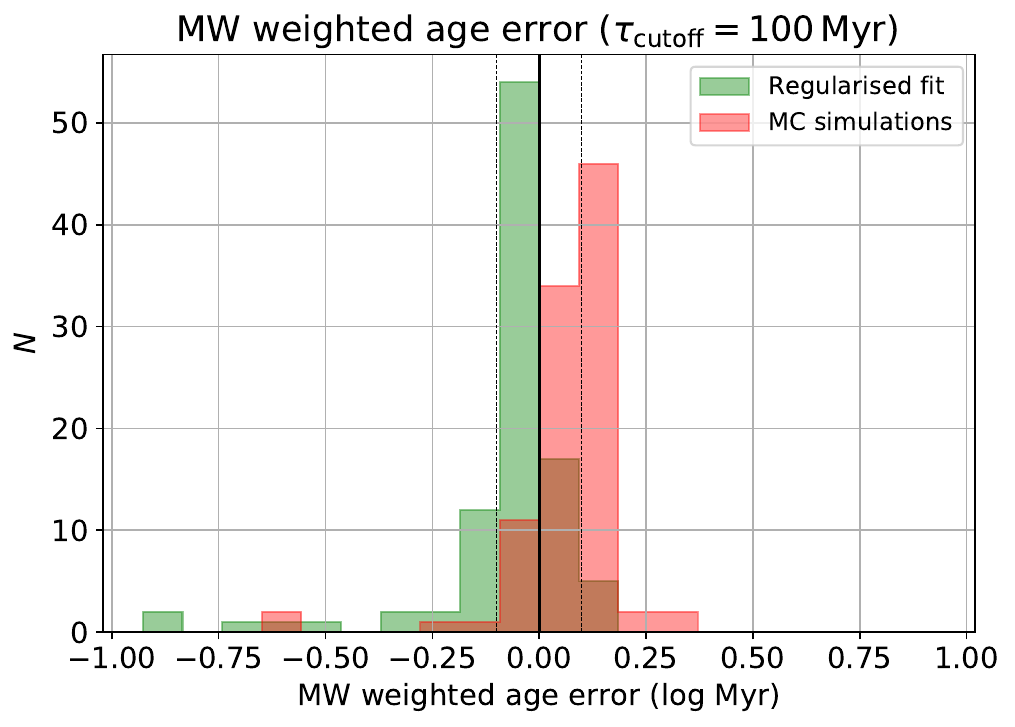}
    \includegraphics[width=0.4\linewidth]{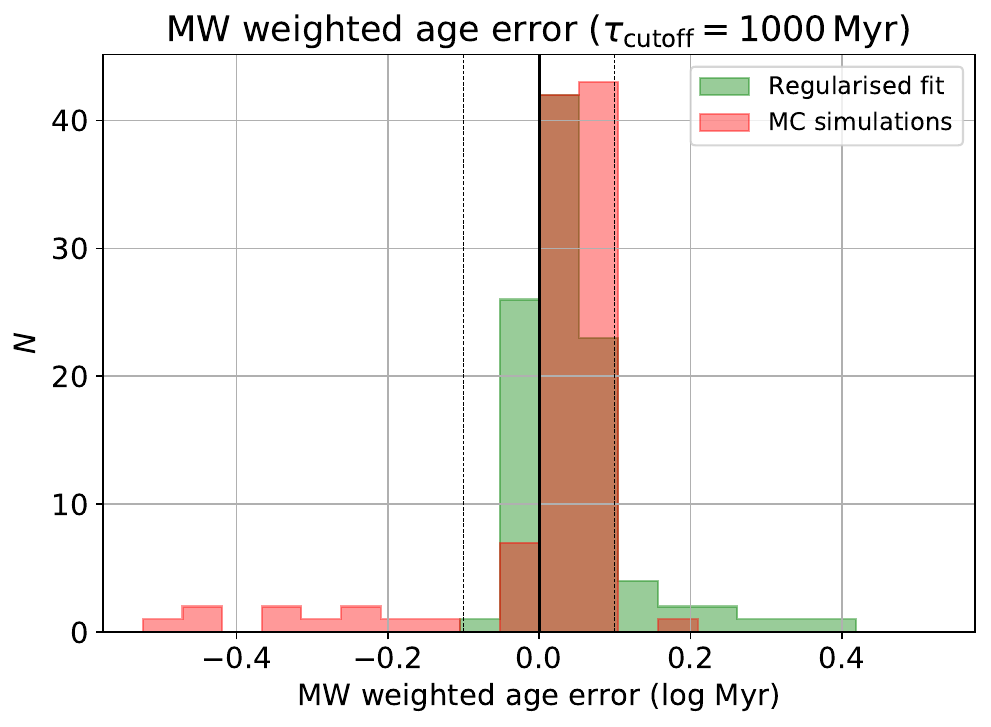}
    \caption{
        Top row: mass-weighted mean ages computed at $\tau_{\rm cutoff} = 100\,\rm Myr$ (left) and $\tau_{\rm cutoff} = 1000\,\rm Myr$ (right) for the 106 most massive galaxies in our mock galaxy sample. The blue, green and red points show the ages computed using the input, regularised and MC best-fit SFHs respectively. The error bars for the MC measurements correspond to the standard deviation of the values computed for all individual runs in the MC simulation.
        Bottom row: histograms showing the distribution in the errors of the computed ages.
    }
    \label{fig: best-case tests - mass-weighted}
\end{figure*}

\begin{figure*}
    \centering
    \includegraphics[width=0.4\linewidth]{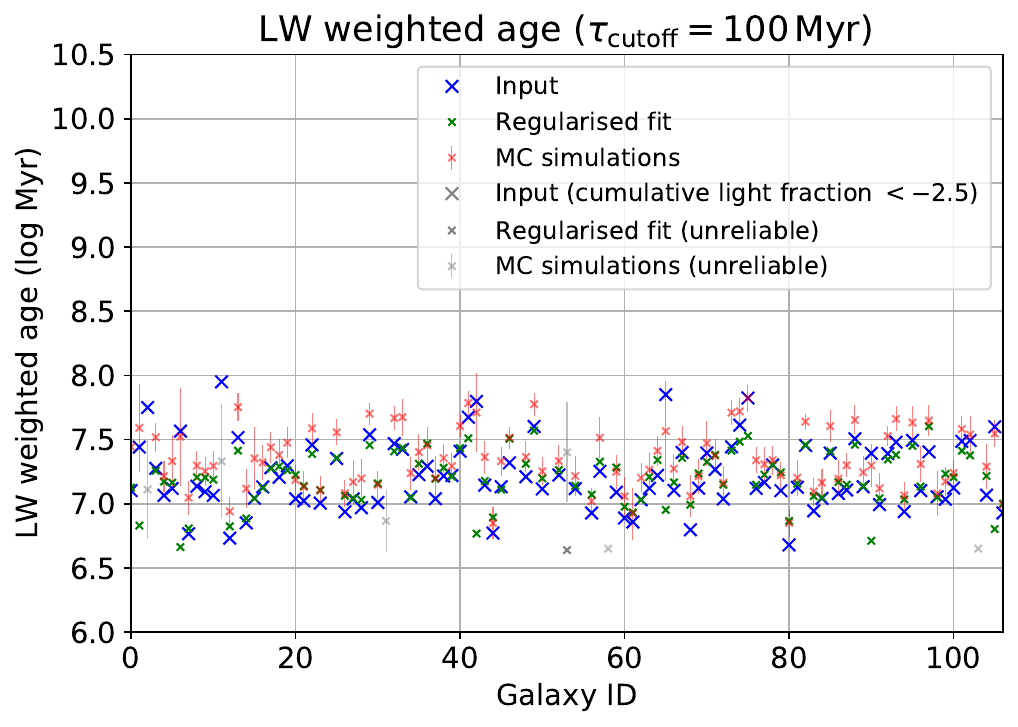}
    \includegraphics[width=0.4\linewidth]{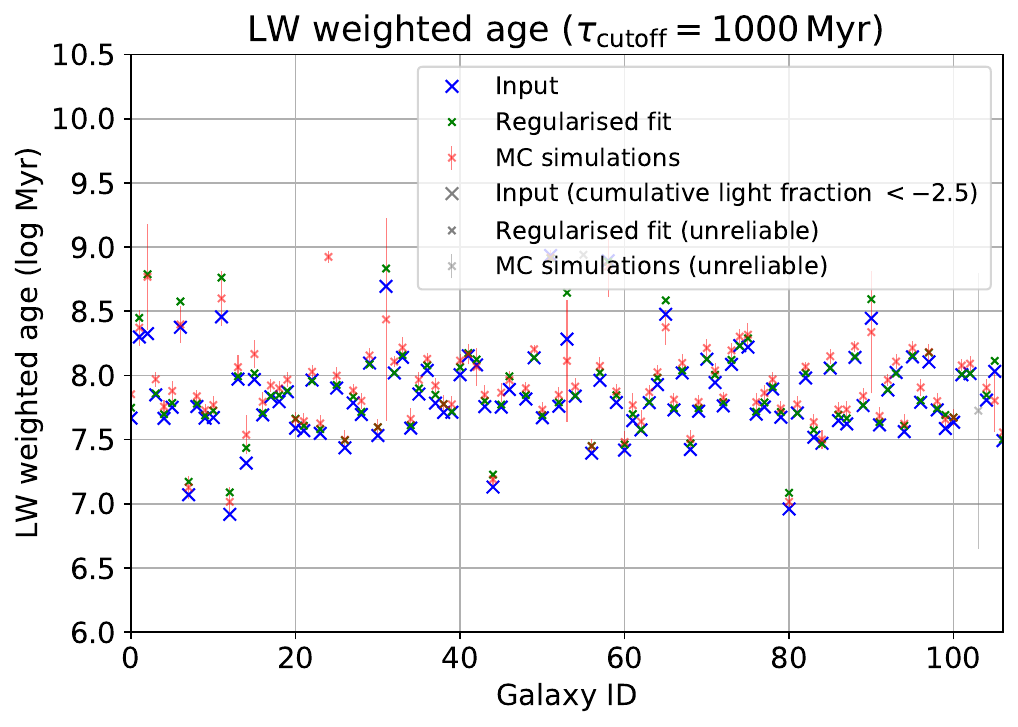}
    \includegraphics[width=0.4\linewidth]{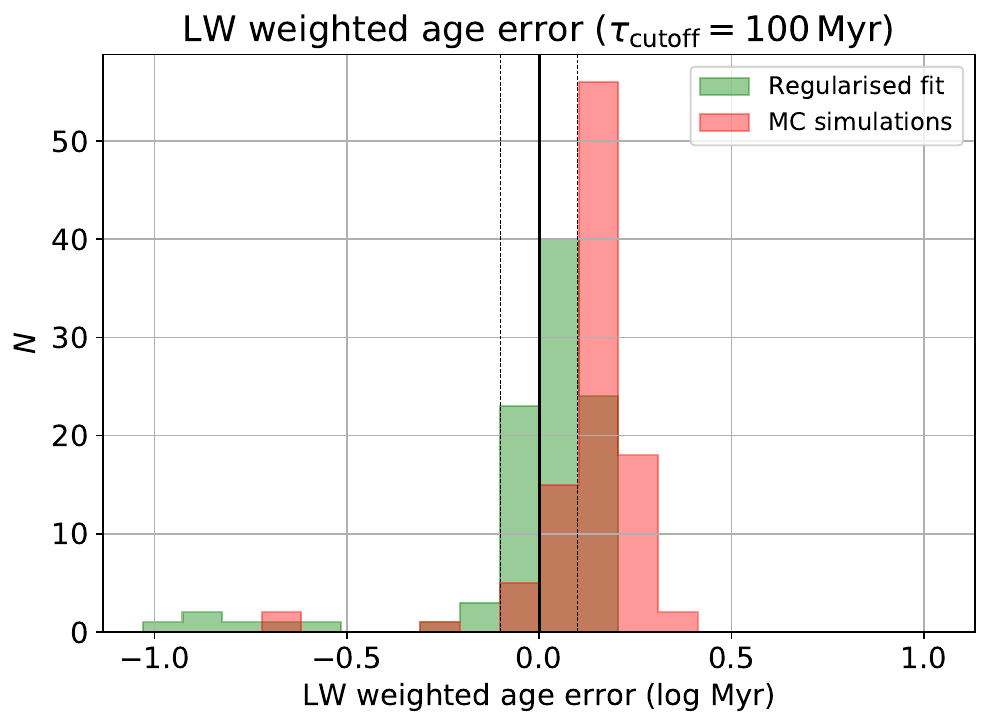}
    \includegraphics[width=0.4\linewidth]{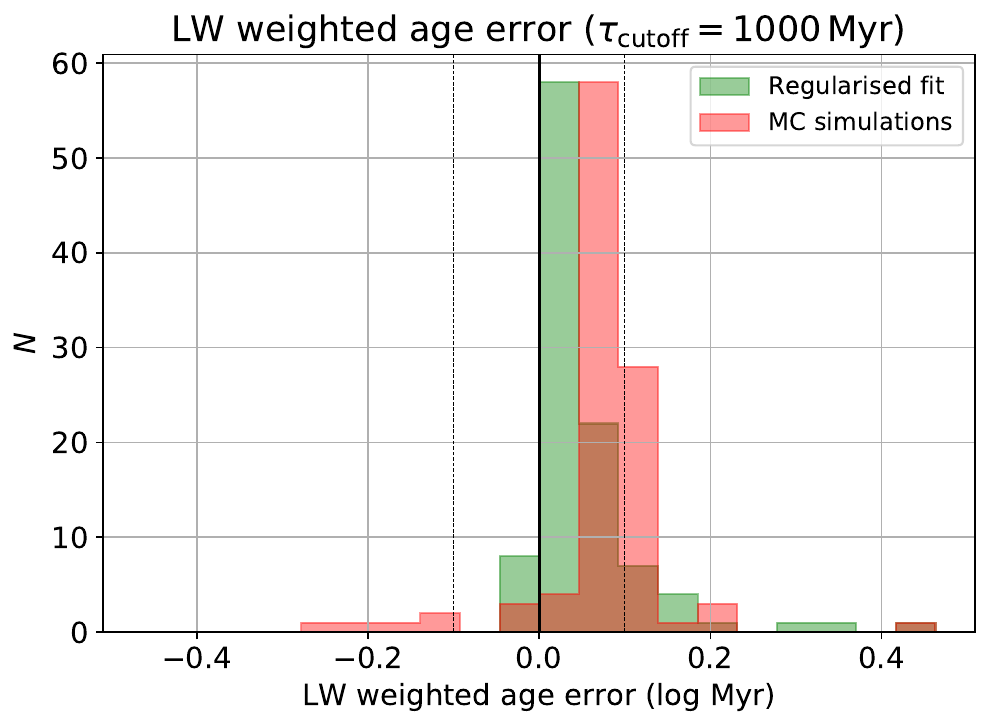}
    \caption{Same as Fig.\ref{fig: best-case tests - mass-weighted} but for light-weighted ages.}
    \label{fig: best-case tests - light-weighted}
\end{figure*}

The top rows of Figs.~\ref{fig: best-case tests - mass-weighted} and \ref{fig: best-case tests - light-weighted} show the MW and LW ages for all 106 galaxies from both the regularised and MC fits, plus the true value measured from the input SFH. 
The LW ages tend to be more sensitive to young stellar populations, and span a larger range than the MW ages; interestingly, ages estimated from the MC fits are systematically higher than those computed from the regularised fits.

The bottom rows show the difference between the measured and true values.
At 100 Myr, the mean error in the mass-weighted ages for the MC and regularised fits are approximately 0.1 (regularised)/0.11 (MC) dex, which reduces to 0.05 (regularised)/0.07 (MC) dex at 1 Gyr.
For the light weighted ages the mean error in the MC and regularised fits are approximately 0.12(regularised)/0.17(MC) dex at 100 Myr and 0.05 (regularised)/0.09 (MC) dex at 1 Gyr.
 These represent a lower limit for the uncertainty on these quantities derived when extinction and/or an AGN continuum is present.

\subsection{``False positive'' analysis}

We found that \ppxf{} often returns solutions with nonzero weights for young stellar templates when no such stellar populations are present in the input. To investigate this issue, we considered ga0024, which has no stellar populations younger than approximately 2 Gyr (see bottom row of ~\autoref{fig: SFHs (all gals)}). 
During our AGN continuum fits (\ref{sec: The impact of extinction and the AGN continuum}), we found that \ppxf{} returned non-zero weights for very young stellar templates when stronger and steeper AGN continua were added, likely due to the degeneracy between the blackbody-like spectra of young stars and steep power-law continua. 
These spurious weights in the young stellar templates rarely exceeded a total light fraction at 4020\,Å of $10^{-3}$. 
In computing our mass- and light-weighted ages, these age estimates were therefore marked as unreliable when the cumulative light fraction below $\tau_{\rm cutoff}$ is $< 10^{-3}$.


\subsection{The impact of extinction and the AGN continuum}\label{sec: The impact of extinction and the AGN continuum}

The tests described in \ref{subsec: appendix: "Best case" scenario} did not include extinction or an AGN continuum. We therefore repeated our simulations with these two factors included in the mock spectra to investigate how they affect the best-fit SFH.

\begin{figure*}
    \includegraphics[width=0.7\linewidth]{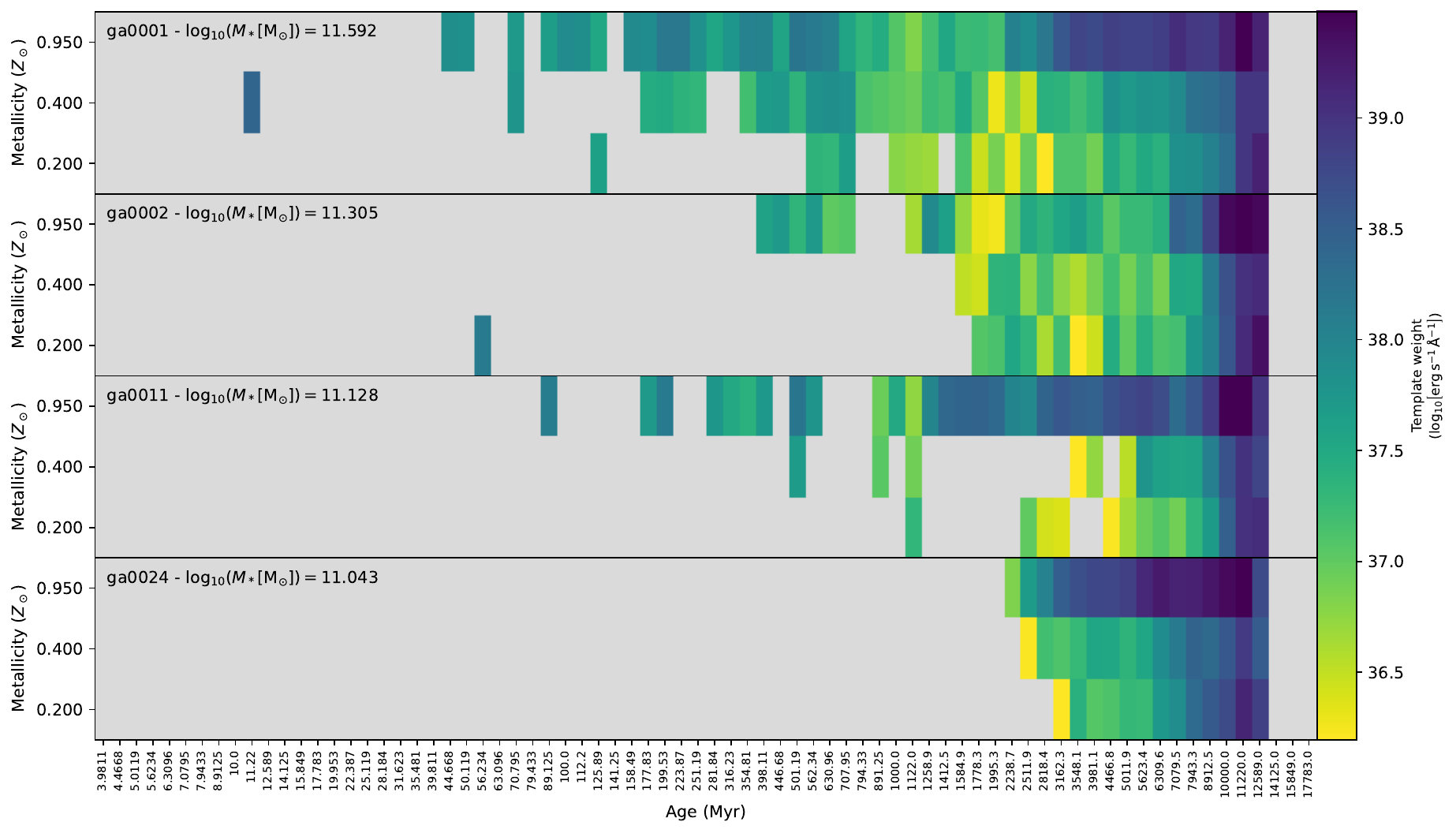}
    \caption{SFHs/CEHs for galaxies ga0001, ga0002, ga0011, and ga0024.}
    \label{fig: SFHs (all gals)}
\end{figure*}

The precise impact that an AGN continuum will have upon the derived results is sensitive to the shape of the input SFH, as shown by \citet{Cardoso2017}, who found that an AGN continuum had different effects upon the best-fit SFHs when the input SFH was characterised by continuous versus instantaneous bursts of star formation. 
As such, we ran our tests on five different galaxies, chosen to have a variety of SFHs:
\begin{itemize}
    \item ga0001 ($M_* = 3.9 \times 10^{11}\msol$): smoothly declining SFH from $\sim 13\,\rm Gyr$ to $\sim 40 \,\rm Myr$ with a recent star formation event ($\sim 10^7\msol$) at 10 Myr.
    \item ga0002 ($M_* = 2.0 \times 10^{11}\msol$): rapidly declining SFH with two bursty star formation events at $\sim 500\,\rm Myr$ and $\sim 60\,\rm Myr$ respectively.
    \item ga0011 ($M_* = 1.3 \times 10^{11}\msol$): smoothly declining SFH with some low-level star formation events between $70 \,\rm  Myr$ and $1\,\rm Gyr$.
    \item ga0024 ($M_* = 1.1 \times 10^{11}\msol$): purely old ($\gtrsim 2\,\rm Gyr$) stellar population with no recent star formation.
    \item ga0044 ($M_* = 3.3 \times 10^{10}\msol$): smoothly declining SFH with a significant ($\sim 10^7\msol$) recent star formation event at $\sim 3 \,\rm Myr$.
\end{itemize}
The SFHs/CEHs for galaxies ga0001, ga0002, ga0011, and ga0024 are shown in ~\autoref{fig: SFHs (all gals)}.

For each galaxy, we created a grid of mock spectra varying in $\alpha_\nu$, $x_{\rm AGN}$ and $A_V$, keeping the S/N fixed at 100. The parameter space explored by our models is as follows: $A_V \in [0, 0.5, 1.0, 2.0, 3.0, 4.0]$, $\alpha_\nu \in [0.5, 1.0, 2.0]$ and $x_{\rm AGN} \in [0.1, 0.2, 0.5, 1.0, 2.0]$, plus models where no AGN continuum was added. \ppxf{} was run on each spectrum using both the regularised and MC methods. 

\subsubsection{Recovery of stellar extinction}
\begin{figure*}
    \centering
    \includegraphics[width=0.8\linewidth]{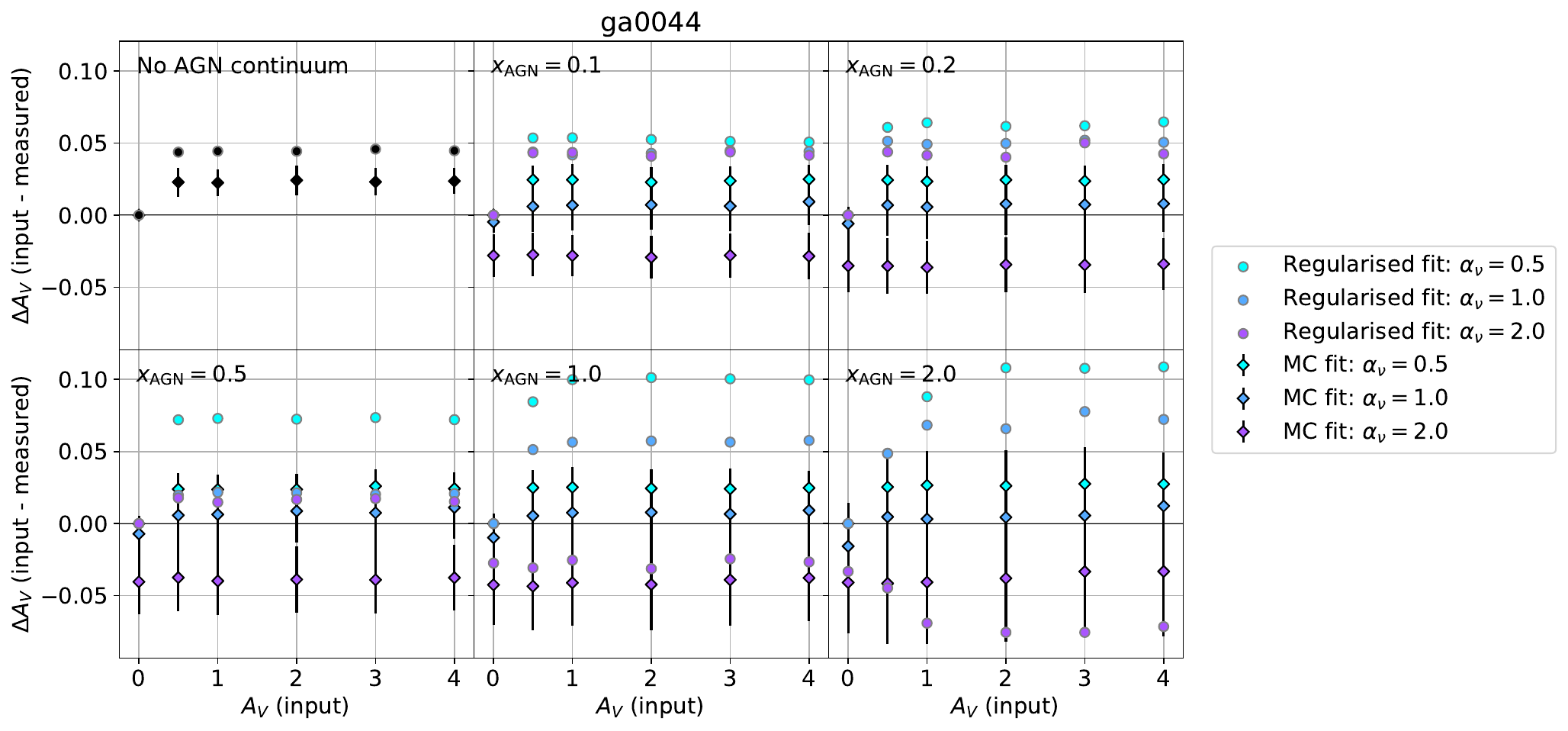}
    \caption{Errors in the best-fit $A_V$ as a function of input $A_V$ for varying AGN continuum parameters $x_{\rm AGN}$ and $\alpha_\nu$.}
    \label{fig: ga0044: A_V recovery}
\end{figure*}

~\autoref{fig: ga0044: A_V recovery} shows the errors in the recovered $A_V$ for ga0044 as a function of both $x_{\rm AGN}$ and $\alpha_\nu$, expressed in both magnitudes and as a percentage. 
For all five of our test galaxies, \ppxf{} was able to recover the input $A_V$ to within $0.1\rm \,mag$ regardless of the presence of an AGN continuum.
When no extinction is included in the input, \ppxf{} correctly returns an $A_V$ close to zero. 
When $A_V > 0$, a stronger and steeper AGN continuum, i.e. a larger $x_{\rm AGN}$ and smaller $\alpha_\nu$, tends to yield larger systematic errors in the $A_V$. 
In particular, \ppxf{} tends to over-estimate $A_V$ when $\alpha_\nu$ is smaller. This is likely due to the partial degeneracy between the extinction curve, which acts to suppress shorter wavelengths, and the AGN continuum, which acts to boost shorter wavelengths when $\alpha_\nu < 1.0$. 

None the less, in all cases, the systematic error in $A_V$ rarely exceeds 0.1\,mag; and this only occurs for the largest values of $x_{\rm AGN}$ which are rare in our sample (see ~\autoref{fig: appendix: S7 ppxf galaxy measurements}). 
We also found the recovered SFH to be essentially independent of the $A_V$ applied; we therefore assumed that stellar extinction does not affect the recovered stellar ages.

\subsubsection{Recovery of the AGN continuum}
As discussed in Section~\ref{subsec: Stellar continuum fitting with ppxf}, the AGN continuum is fit by including four individual templates varying in $\alpha_\nu$ in the \ppxf\ fit. The best-fit $x_{\rm AGN}$ is therefore equal to the total sum of the AGN template weights. 

In general, \ppxf{} accurately recovers $x_{\rm AGN}$. 
When no AGN continuum was included in the input, \ppxf{} rarely returned solutions with $x_{\rm AGN} \gtrsim 0.05$, suggesting that such ``false positives'' are unlikely.
In all of our test galaxies except for ga0044, the error in $x_{\rm AGN}$ rarely exceeds 10\,per\,cent when an AGN continuum is present. In particular, \ppxf{} systematically over-estimates $x_{\rm AGN}$ by up to $\sim 0.1$.

In ga0044, on the other hand, \ppxf{} significantly \textit{under-}estimates $x_{\rm AGN}$.
As shown in ~\autoref{fig: ga0044: SFH/CEH}, ga0044 exhibits a very recent burst of star formation, which represents a significant fraction of the total stellar luminosity at the reference wavelength.
As discussed previously, the blackbody-like spectrum of a very young SSP has a very similar shape to a steep power-law continuum; the resulting degeneracy has likely in this case resulted in an inaccurate fit to the AGN continuum.

\begin{figure*}
    \centering
    \includegraphics[width=0.8\linewidth]{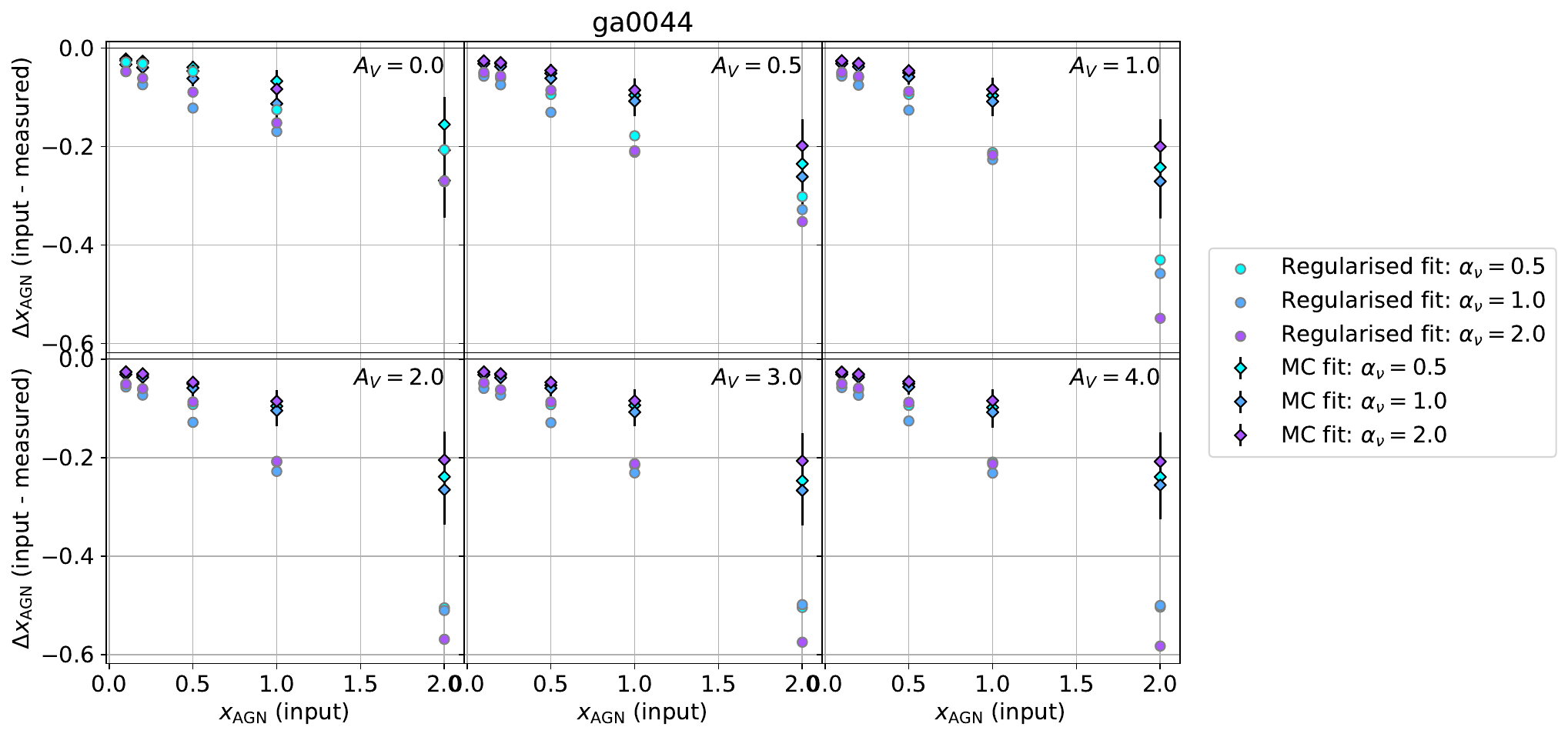}
    \caption{Errors in the best-fit $x_{\rm AGN}$ as a function of input $x_{\rm AGN}$ for varying $A_V$ and $\alpha_\nu$.}
    \label{fig: ga0044: x_AGN recovery}
\end{figure*}

\begin{figure}
    \centering
    \includegraphics[width=1\linewidth]{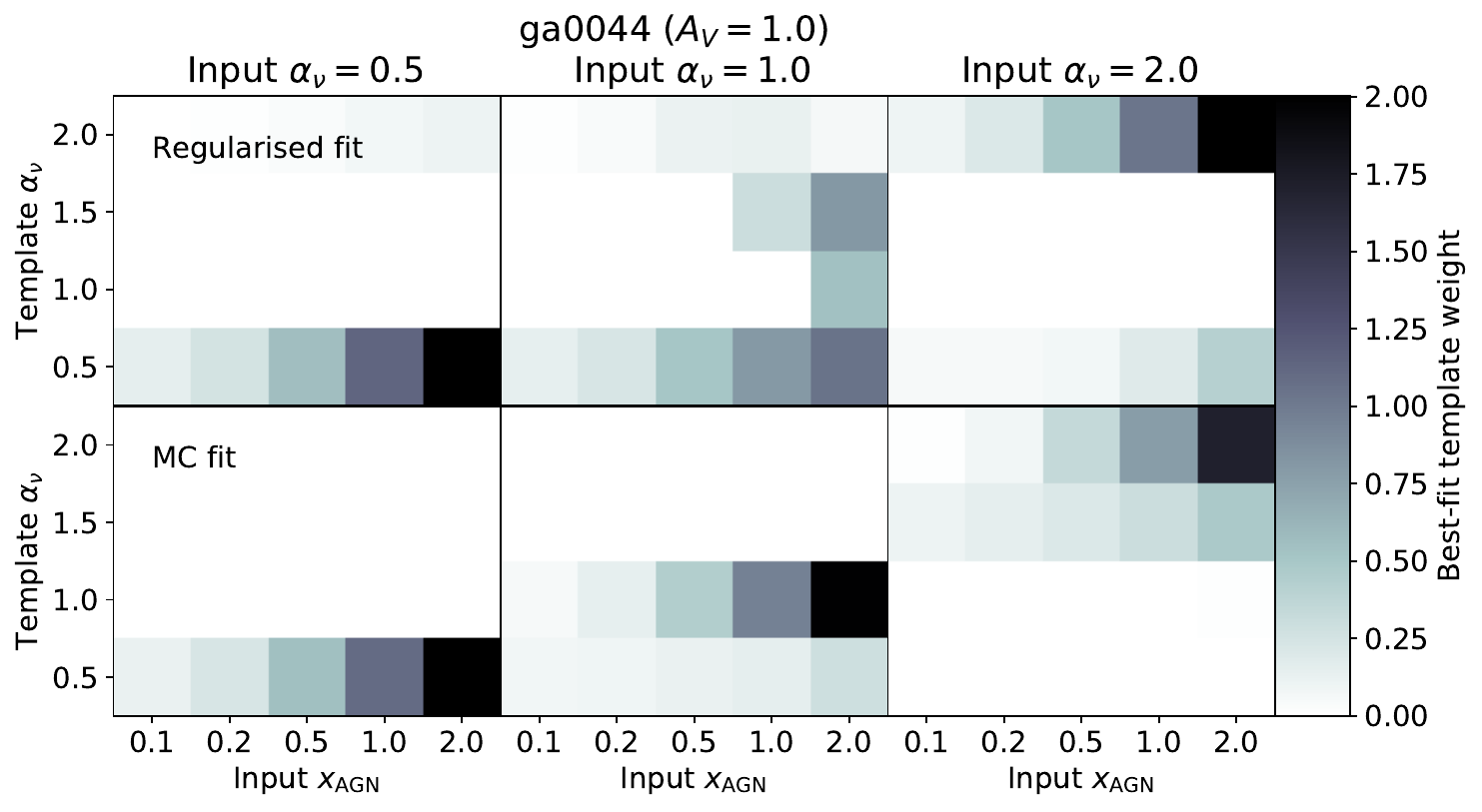}
    \caption{Best-fit AGN template weights as a function of the input $\alpha_\nu$ and $x_{\rm AGN}$ with $A_V = 1.0$ computed using regularisation (top row) and MC simulations (bottom row).}
    \label{fig: ga0044: alpha_nu recovery}
\end{figure}

Because the AGN continuum is fitted in a non-parametric way, we cannot directly recover the best-fit $\alpha_\nu$; however we may still compare the best-fit AGN template weights to the input, as shown in ~\autoref{fig: ga0044: alpha_nu recovery}.
The template with the highest weight in the best-fit solution generally corresponds to the input $\alpha_\nu$, although occasionally a solution will be preferred where there are non-zero weights in several templates, even if the input $\alpha_\nu$ corresponds to that of one of the templates in \ppxf{}; for ga0044 this is most evident in the regularised fit when the input $\alpha_\nu = 1.0$. 

\subsubsection{The impact of the AGN continuum on stellar age determination}\label{subsubsec: appendix: The impact of the AGN continuum}
In most cases, our \ppxf{} implementation is able to accurately recover $x_{\rm AGN}$. We now explore the effect this has upon the derived stellar ages.

\begin{figure*}
    \centering
    \includegraphics[width=0.8\linewidth]{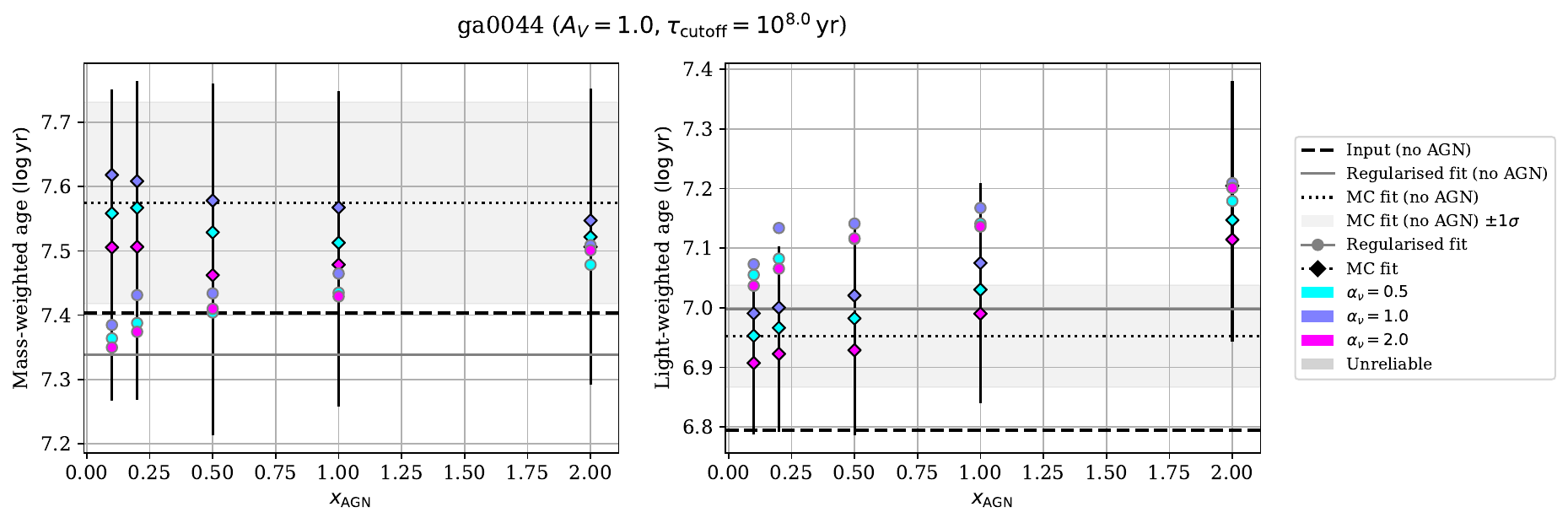}
    \includegraphics[width=0.8\linewidth]{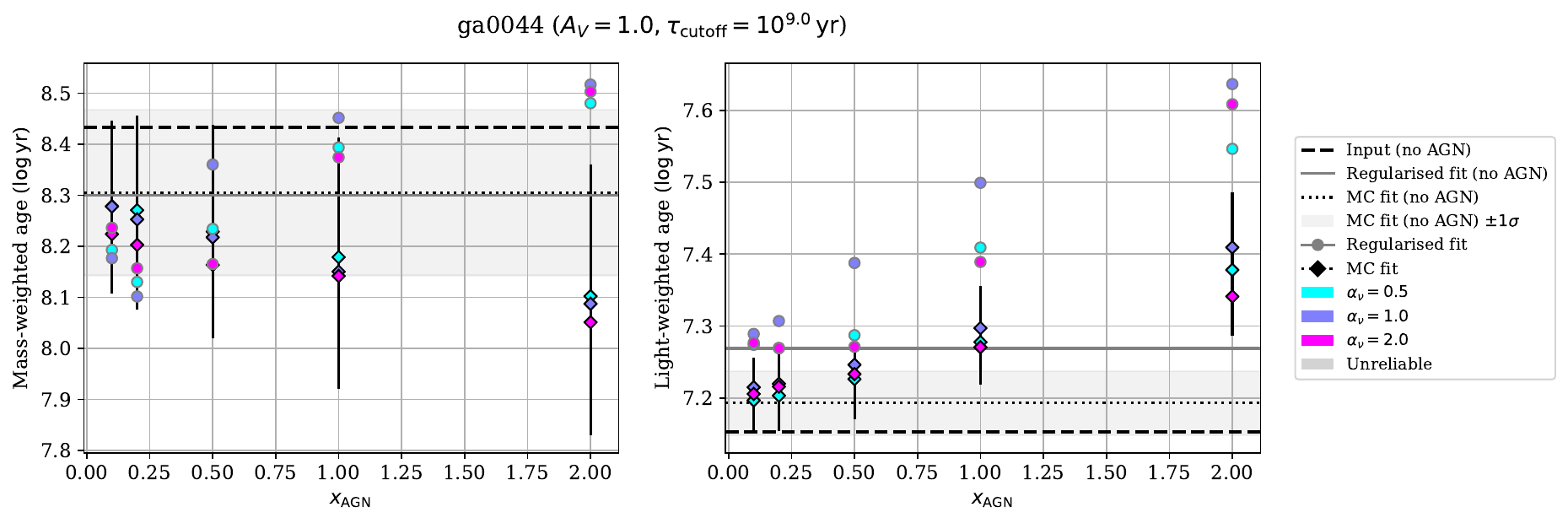}
    \caption{Mass-weighted (left panels) and light-weighted (right panels) ages computed for ga0044 with $\tau_{\rm cutoff} = 10^8\,\rm yr$ (top row) and $\tau_{\rm cutoff} = 10^9\,\rm yr$ (bottom row)  as a function of both $x_{\rm AGN}$ and $\alpha_\nu$. The dashed horizontal line shows the true value, the solid grey line shows the value recovered using the regularised approach without an AGN continuum, and the black dotted line shows the mean value computed from the MC simulations without an AGN continuum, with the grey shaded region indicating the $\pm1\sigma$ range.}
    \label{fig: ga0044: MW/LW ages vs. AGN parameters}
\end{figure*}

In ~\autoref{fig: ga0044: MW/LW ages vs. AGN parameters} we show the mass- and light-weighted ages computed below 100 Myr and 1 Gyr as a function of $x_{\rm AGN}$ and $\alpha_\nu$ for ga0044. For the sake of brevity, we only show results for $A_V = 1.0\,\rm mag$, as the results are similar regardless of the value of $A_V$.
For this particular galaxy, when $\tau_{\rm cutoff} = 10^8\,\rm yr$, the systematic error on the mass-weighted ages are comparable to those when no AGN continuum is present; this is true for both the MC and regularised fits. There are some trends present: as $x_{\rm AGN}$ increases, so too does the error in the regularised fit, whereas the errors in the MC fit generally decrease with $x_{\rm AGN}$.
The errors in the light-weighted ages, on the other hand, increase substantially as a function of $x_{\rm AGN}$, reading values as high as 0.4 dex when $x_{\rm AGN} = 2.0$. This is true for both values of $\tau_{\rm cutoff}$ assumed.
To summarise, for ga0044, the errors \textit{generally} increase with $x_{\rm AGN}$, such that the systematic error in the LW and MW ages can be as high as 0.4 dex. But, the error is considerably smaller when $x_{\rm AGN}$ is small. 

Analysis of results from the other four galaxies show that firstly, the magnitude of the errors in the MW/LW ages are proportional to $x_{\rm AGN}$: the larger $x_{\rm AGN}$, the larger are the systematic errors in the recovered age. Secondly, the recovered ages generally \textit{increase} with $x_{\rm AGN}$, but not in all cases. Similarly, in general, the lower the $\alpha_\nu$, i.e. the steeper the AGN continuum, the \textit{lower} the estimated age, but not in all cases.

\subsection{Conclusions}\label{subsec: appendix: Conclusions}

We draw the following conclusions from these tests on mock spectra:

\begin{itemize}
    \item When no AGN continuum is present, the mean absolute error in the LW/MW ages is $\sim 0.1\,\rm dex$ when $\tau_{\rm cutoff} = 1\,\rm Gyr$ and $\sim 0.2\,\rm dex$ when $\tau_{\rm cutoff} = 100\,\rm Myr$. These values represent the minimum systematic errors we should apply to our age measurements.
    \item Both the MC and regularised fits yield solutions with similar MW/LW ages, although they often disagree in these values. Which of the MC or regularised fits yields a more accurate answer most likely depends upon the intrinsic shape of the galaxy's SFH: for those with smoother SFHs, the regularised fit is likely to be more accurate, whereas for those with spikier SFHs the MC result may be more accurate. Unfortunately as the "spikiness" of the SFH is not known a-priori for any of our galaxies, we are unable to predict which of the two methods will produce the most accurate result. However, the MC method provides uncertainties on the SFH, unlike the regularised method, and is less prone to numerical instabilities. We therefore use only the MC fits for our analysis.
    \item \ppxf{} rarely reports the presence of an AGN continuum when none exists in the input. We therefore assume that no optically-significant AGN continuum is present when our \ppxf{} results return $x_{\rm AGN} \sim 0$.
    \item Typical errors in the retrieval of $A_V$ are $<0.1\,\rm mag$. We therefore assume the $A_V$ estimates are reliable.
    \item The shape of the best-fit AGN continuum is generally recovered accurately as long as the input value is within the $\alpha_\nu$ range of templates used in \ppxf{}.
    \item The systematic errors in LW ages scale with $x_{\rm AGN}$, reaching 0.4\,dex when there is significant contamination from an AGN continuum. However, given that we can usually accurately estimate $x_{\rm AGN}$, we can estimate an appropriate error in the LW age using the best-fit $x_{\rm AGN}$. We therefore recommend assigning systematic errors based on the recovered $x_{\rm AGN}$: if $x_{\rm AGN} \sim 0$ then we can use our estimates from the "best-case" fits. If the $x_{\rm AGN} \sim 1$, an additional systematic error of 0.4\,dex should be added to the uncertainty.
\end{itemize}

%% file: ppxf_extra_figs.tex
\begin{figure}
    \centering
    \includegraphics[width=1\linewidth]{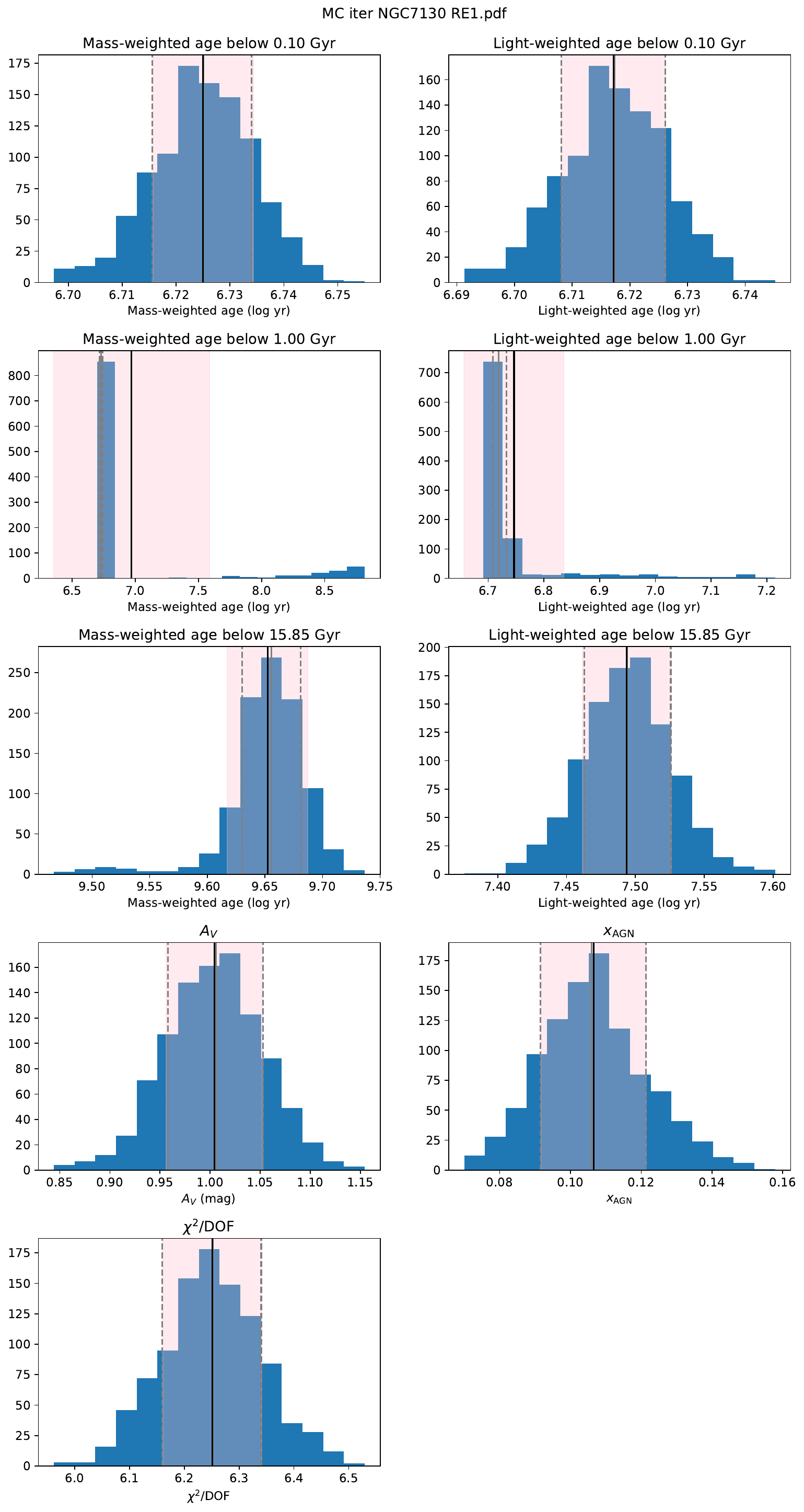}
    \caption{Distributions in various parameters resulting from the 1000 \ppxf\ MC fits to the $1\re$ spectrum of NGC7130. The black vertical line represents the mean, and the pink shaded region represents the mean $\pm 1 \sigma$. The 50th percentile is represented by the solid grey line, and the dashed grey lines represent the 16th and 84th percentiles.}
    \label{fig: appendix: NGC7130 MC distributions}
\end{figure}

\begin{figure*}
    \centering
    \includegraphics[width=\linewidth]{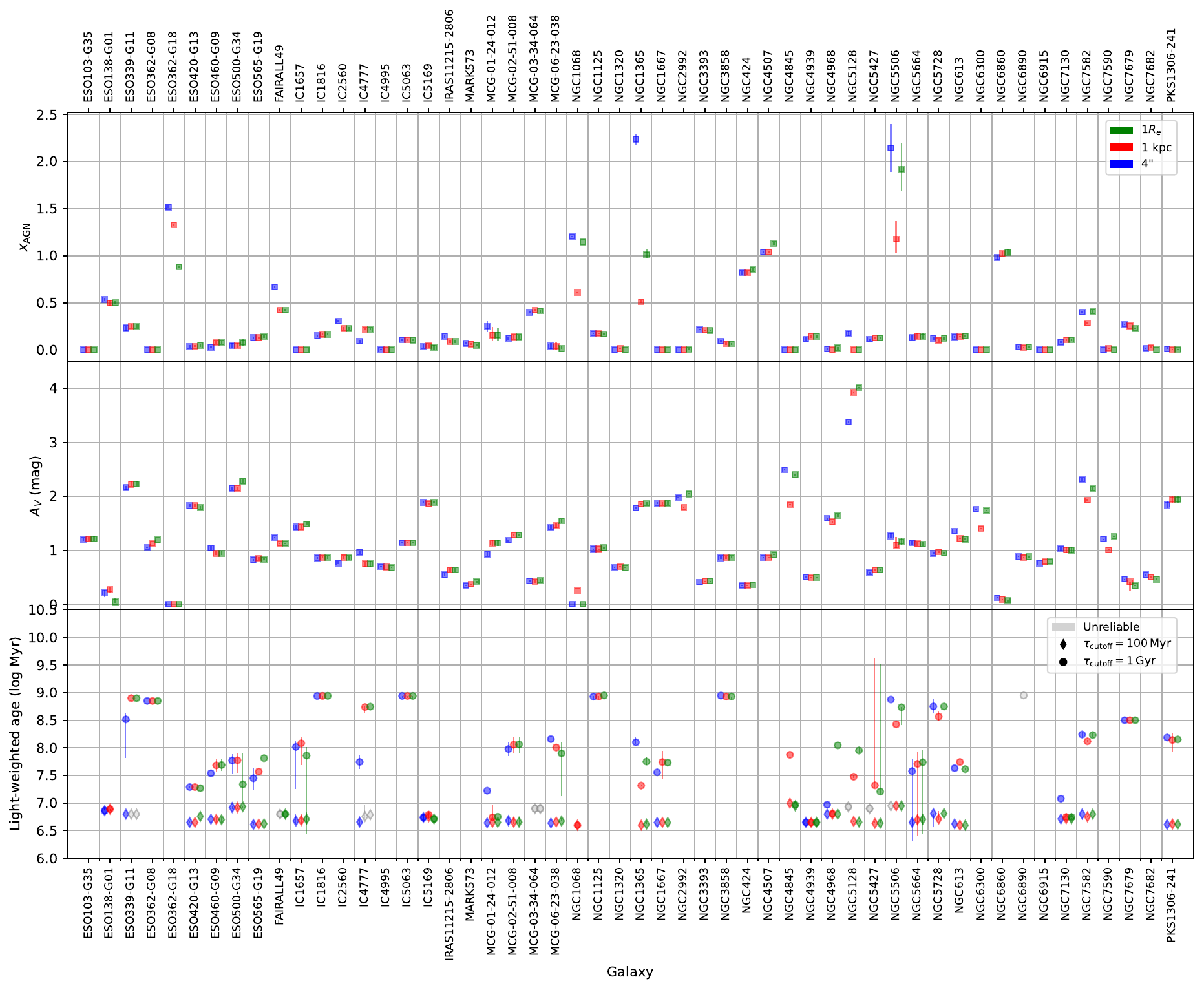}
    \caption{Measurements of $x_{\rm AGN}$ (top), stellar $A_V$ (middle) and LW ages (bottom) for the galaxies in our sample using the methods described in Section~\ref{subsec: appendix: ppxf fitting method} using \ppxf{}. Blue, red and green points represent measurements from the $1\re$, 1\,kpc and 4'' aperture spectra respectively, and the vertical error bars represent the 16th and 84th percentile confidence intervals. Points that are missing correspond to instances where there were no non-zero template weights below the corresponding age cutoff.}
    \label{fig: appendix: S7 ppxf galaxy measurements}
\end{figure*}

Measurements of $x_{\rm AGN}$ (top), stellar $A_V$ (middle) and LW ages (bottom) for the galaxies in our sample using the methods described in~\autoref{sec: Stellar age determination} are shown in~\autoref{fig: appendix: S7 ppxf galaxy measurements}.

